\documentclass[structabstract]{aa} 
\usepackage{longtable}
\usepackage{graphicx}
\usepackage{txfonts}
\usepackage{natbib}

\begin{document}
  \title{New effective recombination coefficients for nebular \ion{N}{ii} lines
     \thanks{Complete Tables~\ref{aeff_Ne100} -- \ref{aeff_Ne100000} in 
             electronic form only are available at the CDS via anonymous ftp 
             to cdarc.u-strasbg.fr (130.79.128.5) or via 
             http://cdsweb.u-strasbg.fr/Abstract.html.}}
  \author{X. Fang \inst{1}, P. J. Storey \inst{2} \and 
          X.-W. Liu \inst{1,}\inst{3}}
  \offprints{X. Fang}
  \institute{Department of Astronomy, School of Physics, Peking University, 
             Beijing 100871, P.~R.~China\\
             \email{fangx@vega.bac.pku.edu.cn}
             \and
             Department of Physics and Astronomy, University College London, 
             Gower Street, London WC1E 6BT, UK
             \and
             The Kavli Institute for Astronomy and Astrophysics, Peking 
             University, Beijing 100871, P.~R.~China}

  \date{Received / Accepted }

\abstract {}
{In nebular astrophysics, there has been a long-standing dichotomy in plasma 
diagnostics between abundance determinations using the traditional method 
based on collisionally excited lines (CELs), on the one, hand and (optical) 
recombination lines/continuum, on the other. A number of mechanisms have been 
proposed to explain the dichotomy. Deep spectroscopy and recombination line 
analysis of emission line nebulae (planetary nebulae and \ion{H}{ii} regions) 
in the past decade have pointed to the existence of another previously unknown
component of cold, \ion{H}{}-deficient material as the culprit. Better 
constraints are needed on the physical conditions, chemical composition, 
mass, and spatial distribution of the postulated \ion{H}{}-deficient inclusions
in order to unravel their astrophysical origins. This requires knowledge of the
relevant atomic parameters, most importantly the effective recombination 
coefficients of abundant heavy element ions such as \ion{C}{ii}, \ion{O}{ii}, 
\ion{N}{ii}, and \ion{Ne}{ii}, appropriate for the physical conditions 
prevailing in those cold inclusions (e.g. $T_\mathrm{e} \leq 1,000$~K ).}
{Here we report new ${\it ab~initio}$ calculations of the effective 
recombination coefficients for the \ion{N}{ii} recombination spectrum. We 
have taken into account the density dependence of the coefficients arising 
from the relative populations of the fine-structure levels of the 
ground state of the recombining ion, an elaboration that has not been 
attempted before for this ion, and it opens up 
the possibility of electron density determination via 
recombination line analysis. Photoionization cross-sections, bound state 
energies, and the oscillator strengths of \ion{N}{ii} with $n \leq 11$ and 
$l \leq 4$ have been obtained using the close-coupling R-matrix method in 
the intermediate coupling scheme. Photoionization data were computed
that accurately map out the near-threshold resonances and 
were used to derive recombination coefficients, including radiative 
and dielectronic recombination. Also new is including the effects of 
dielectronic recombination via high-$n$ resonances lying between the 
$^2$P$^{\rm o}$\,$_{1/2}$ and $^2$P$^{\rm o}$\,$_{3/2}$ levels. The new 
calculations are valid for temperatures down to an unprecedentedly low level
(approximately $100$~K). The newly calculated effective recombination 
coefficients allow us to construct plasma diagnostics based on the measured 
strengths of the \ion{N}{ii} optical recombination lines (ORLs).}
 {The derived effective recombination coefficients are fitted with analytic 
formulae as a function of electron temperature for different electron 
densities. The dependence of the emissivities of the strongest transitions 
of \ion{N}{ii} on electron density and temperature is illustrated. Potential 
applications of the current data to electron density and temperature 
diagnostics for photoionized gaseous nebulae are discussed. We also present 
a method of determining electron temperature and density 
simultaneously.} {}
\keywords{atomic data -- line: formation -- \ion{H}{ii} regions -- ISM: atoms --
          planetary nebulae: general}

\titlerunning{Effective Recombination Coefficients for \ion{N}{ii} Lines}
\authorrunning{X.~Fang et al.}

\maketitle

\section{Introduction}

The principal means of electron temperature and density diagnostics
and heavy-element abundance determinations in nebular plasmas 
has, until recently, been measurement of collisionally excited lines (CELs). 
The emissivities of CELs have, however, an exponential dependence on electron 
temperature and, consequently, so do the heavy element abundances deduced 
from them. An alternative method of determining heavy element abundances is 
to divide the intensities of ORLs emitted by heavy element ions with 
those by hydrogen. Such ratios are only weakly dependent on temperature. In 
planetary nebulae (PNe), abundances of \ion{C}{}, \ion{N}{}, and \ion{O}{} 
derived from ORLs have been shown to be systematically larger than those 
derived from CELs 
typically by a factor of 2. Discrepancies of much larger magnitudes, say by 
more than a factor of 5, are also found for a small fraction of PNe (about 
10\%; Liu et al. \citealt{Liu95}, \citealt{Liu2000}, \citealt{Liu2001}, 
\citealt{Liu2006b}; Liu et al. \citealt{Liu04}; Luo et al. \citealt{Luo01}). 
In the most extreme case, the abundance discrepancy factor (ADF) reaches a 
record value of 70 (Liu et al. \citealt{lbzbs2006}). There is strong evidence
that nebulae contain another component of metal-rich, cold plasma, probably
in the form of \ion{H}{}-deficient inclusions embedded in the diffuse
gas (Liu et al. \citealt{Liu2000}). The existence of cold inclusions
provides a natural explanation to the long-standing dichotomy of
abundance determinations and plasma diagnostics (Liu \citealt{Liu03},
\citealt{Liu2006a}, \citealt{Liu2006b}). The prerequisite for reliable
determinations of recombination line abundances is accurate effective
recombination coefficients for heavy element recombination lines. In
this paper, we present new effective recombination coefficients for
the recombination spectrum of \ion{N}{ii}.

Radiative recombination coefficients of \ion{N}{ii} have been given by 
P\'{e}quignot et al. \cite{Pequignot91}. Nussbaumer \& Storey \cite{NS84} 
tabulate dielectronic recombination coefficients of \ion{N}{ii} obtained from
a model in which resonance states are represented by bound-state wave 
functions. Escalante \& Victor \cite{EV90} calculate effective recombination 
coefficients for \ion{C}{i} and \ion{N}{ii} lines using an atomic model 
potential approximation for transition probabilities and recombination 
cross-sections. They then add the contribution from dielectronic recombination
using the results of Nussbaumer \& Storey \cite{NS84}.

In the most recent work of \ion{N}{ii} by Kisielius \& Storey
\cite{KS02}, they follow the approach of Storey \cite{Storey94} who,
in dealing with \ion{O}{ii}, uses a unified method for the treatment of
radiative and dielectronic recombination by directly calculating
recombination coefficients from photoionization cross-sections of each
initial state. They also incorporate improvements introduced by
Kisielius et al. \cite{KS98} in their work on \ion{Ne}{ii}. The
\ion{N}{$^+$} photoionization cross-sections were calculated in 
$LS$-coupling using the {\it ab initio} methods developed for the
Opacity Project (Seaton \citealt{Seaton87}; Berrington et
al. \citealt{OP1}) and the Iron Project (Hummer et al. \citealt{IP}),
hereafter referred to as the OP methods. The calculations employed the
R-matrix formulation of the close-coupling method, and the resultant
cross-sections are of higher quality. 
In the photoionization calculations of
Kisielius \& Storey \cite{KS02}, for the energy ranges in which the
resonances make main contributions to the total recombination, an
adaptive energy mesh was used to map out all the strong resonances
near thresholds. This approach has also been adopted in our current
calculations, and it will be described in more details in a later
section. In the work of Kisielius \& Storey \cite{KS02}, transition
probabilities for all low-lying bound states were also calculated
using the close-coupling method, so that the bound-bound and
bound-free radiative data used for calculating the recombination
coefficients formed a self-consistent set of data and were expected to
be significantly more accurate than those employed in earlier work.

Here we report new calculations of the effective recombination
coefficients for the \ion{N}{ii} recombination spectrum. Hitherto few
high quality atomic data were available to diagnose plasmas of 
very low temperatures ($\leq 1000$~K), such as the cold, H-deficient
inclusions postulated to exist in PNe (Liu et
al. \citealt{Liu2000}). We have calculated the effective recombination
coefficients for the \ion{N}{$^+$} ion down to an unprecedentedly low
temperature (about $100$~K). At such low temperatures, dielectronic
recombination via high-$n$ resonances between the \ion{N}{$^{2+}$}
ground $^2$P$^{\rm o}$\,$_{1/2}$ and $^2$P$^{\rm o}$\,$_{3/2}$
fine-structure levels contributes significantly to the total
recombination coefficient. We include such effects in our
calculations. We also take into account the density dependence of the
coefficients through the level populations of the fine-structure
levels of the ground state of the recombining ion 
($^2$P$^{\rm o}$\,$_{1/2,3/2}$ in the case of N$^{2+}$). That 
opens up the possibility of electron density determinations via 
recombination line analysis. With the exception of the calculations 
of Kisielius \& Storey \cite{KS99} on \ion{O}{iii} recombination lines, 
all previous work on nebular recombination lines has been in $LS$-coupling 
and therefore it tacitly assumed that the levels of the ground state are 
populated in proportion to their statistical weights. Photoionization
cross-sections, bound state energies and oscillator strengths of
\ion{N}{ii} with $n \leq 11$ and $l \leq 4$ have been obtained using
the R-matrix method in the intermediate coupling scheme.
The photoionization data are used to derive recombination coefficients, 
including contributions from radiative and dielectronic recombination. 
The results are applicable to PNe, H~{\sc ii} regions and nova shells for 
a wide range of electron temperature and density.

\section{\label{s:data}
Atomic data for \ion{N}{$^+$}}

\subsection{\label{s:data0}
The \ion{N}{$^+$} term scheme}

The principal series of \ion{N}{ii} is 2s$^2$2p($^2$P$^{\rm o}$){\it nl}, 
which gives rise to singlet and triplet terms. Also interspersed are
members of the series 2s2p$^2$($^4$P){\it nl} ({\it n} = 3, 4) giving rise 
to triplet and quintet terms. Higher members of this series lie above the 
first ionization limit, and hence may give rise to low-temperature 
dielectronic recombination. There are a few members of the 
2s2p$^2$($^2$D){\it nl} and 2s2p$^2$($^2$S){\it nl} series located above the 
first ionization threshold which also give rise to resonance structures in 
the photoionization cross-sections for singlets and triplets. For photon 
energies above the second ionization threshold, the main resonance structures 
are due to the 2s2p$^2$($^2$D){\it nl} series with some interlopers from the
2s2p$^2$($^2$S){\it nl} and 2s2p$^2$($^2$P){\it nl} series (Kisielius
\& Storey \citealt{KS02}).

\subsection{\label{s:data1}
New R-matrix calculation}

We have carried out a new calculation of bound state energies, oscillator 
strengths and photoionization cross-sections for \ion{N}{ii} states with 
$n \leq 11$ using the OP methods (Hummer et al. \citealt{IP}).  The 
\ion{N}{$^{2+}$} target configuration set was generated with the general 
purpose atomic structure code SUPERSTRUCTURE (Eissner et al. 
\citealt{Eissner74}) with modifications of Nussbaumer \& Storey \cite{NS78}. 
The target radial wave functions of \ion{N}{$^{2+}$} were then generated 
with another atomic structure code AUTOSTRUCTURE\footnote{AUTOSTRUCTURE is 
developed by the Department of Physics at the University of Strathclyde, 
Glasgow, Scotland. The code is available from the website 
http://amdpp.phys.strath.ac.uk/autos .}, which, developed from SUPERSTRUCTURE
and capable of treating collisions, is able to calculate autoionization rates,
photoionization cross-sections, etc. The original theory of AUTOSTRUCTURE is 
described by Badnell \cite{Badnell86}. The wave functions of the nineteen 
target terms were expanded in terms of the 21 electron configurations
1s$^2$2s$^2$2p, 
1s$^2$2s2p$^2$, 
1s$^2$2p$^3$, 
1s$^2$2s$^2{\bar 3}$s, 
1s$^2$2s$^2{\bar 3}$p, 
1s$^2$2s$^2{\bar 3}$d, 
1s$^2$2s2p${\bar 3}$s, 
1s$^2$2s2p${\bar 3}$p, 
1s$^2$2s2p${\bar 3}$d, 
1s$^2$2p$^2{\bar 3}$s, 
1s$^2$2p$^2{\bar 3}$p, 
1s$^2$2p$^2{\bar 3}$d, 
1s$^2$2s${\bar 3}$d$^2$, 
1s$^2$2p${\bar 3}$d$^2$, 
1s$^2$2s$^2{\bar 4}$s, 
1s$^2$2s$^2{\bar 4}$d, 
1s$^2$2s2p${\bar 4}$s, 
1s$^2$2s2p${\bar 4}$p, 
1s$^2$2s2p${\bar 4}$d, 
1s$^2$2s2p${\bar 4}$f, 
1s$^2$2p$^2{\bar 4}$p, 
where 1s, 2s and 2p are spectroscopic orbitals and ${\bar 3}l$ and
${\bar 4}l^{\prime}$ ($l$ = 0, 1, 2 and $l^{\prime}$ = 0, 1, 2, 3) are
correlation orbitals.  The one-electron radial functions for the 1s, 2s and 
2p orbitals were calculated in adjustable Thomas-Fermi potentials, while 
the radial functions for the remaining orbitals were calculated in Coulomb
potentials of variable nuclear charge, $Z_{nl} = 7|\lambda_{nl}|$. 
The potential scaling parameters $\lambda_{nl}$ were determined by minimizing 
the sum of the energies of the eight energetically lowest target states in 
our model.  We obtained for the potential scaling parameters:
$\lambda_\mathrm{1s} = 1.4279$,
$\lambda_\mathrm{2s} = 1.2840$,
$\lambda_\mathrm{2p} = 1.1818$,
$\lambda_\mathrm{{\bar 3} s} = -0.9696$,
$\lambda_\mathrm{{\bar 3} p} = -0.9054$,
$\lambda_\mathrm{{\bar 3} d} = -1.0924$,
$\lambda_\mathrm{{\bar 4} s} = -1.3871$,
$\lambda_\mathrm{{\bar 4} p} = -1.3821$,
$\lambda_\mathrm{{\bar 4} d} = -1.5797$ and 
$\lambda_\mathrm{{\bar 4} f} = -1.9837$.

\begin{table}
\caption{\label{levels}
Comparison of energies (in Ry) for the \ion{N}{$^{2+}$} target states}
\begin{flushleft}
\begin{tabular}{llrrr} \cline{1-5} \\
\multicolumn{1}{c}{Configuration}&
\multicolumn{1}{c}{Term}&
\multicolumn{1}{c}{present}&
\multicolumn{1}{c}{KS2002$^a$} &
\multicolumn{1}{c}{Experimental$^b$}\\
\cline{1-5} \\
1s$^2$2s$^2$2p         & $^2$P$^\mathrm{o}$ & 0.000000 & 0.00000 & 0.00000 \\
1s$^2$2s2p$^2$         & $^4$P              & 0.509051 & 0.51771 & 0.52091 \\
                       & $^2$D              & 0.933243 & 0.92246 & 0.91960 \\
                       & $^2$S              & 1.235909 & 1.20597 & 1.19279 \\
                       & $^2$P              & 1.362871 & 1.33866 & 1.32898 \\
1s$^2$2p$^3$           & $^4$S$^\mathrm{o}$ & 1.706032 & 1.70403 & 1.70123 \\
                       & $^2$D$^\mathrm{o}$ & 1.872797 & 1.85796 & 1.84962 \\
                       & $^2$P$^\mathrm{o}$ & 2.144377 & 2.11771 & 2.09865 \\
\cline{1-5} \\
\end{tabular}
\end{flushleft}
\begin{description}
\item [$^a$] Theoretical calculations by Kisielius \& Storey \cite{KS02}.
\item [$^b$] Eriksson \cite{Eriksson83}.
\end{description}
\end{table}

\begin{table}
\centering
\caption{\label{targets}
The \ion{N}{$^{2+}$} target terms}
\begin{tabular}{rl} 
\cline{1-2} \\
\multicolumn{1}{c}{Configuration} & \multicolumn{1}{c}{Term}\\
\cline{1-2} \\
1s$^2$2s$^2$2p                            & $^2$P$^\mathrm{o}$\\
1s$^2$2s2p$^2$                            & $^4$P\\
                                          & $^2$D\\
                                          & $^2$S\\
                                          & $^2$P\\
1s$^2$2p$^3$                              & $^4$S$^\mathrm{o}$\\
                                          & $^2$D$^\mathrm{o}$\\
                                          & $^2$P$^\mathrm{o}$\\
1s$^2$2s$^2{\bar 3}$s                     & $^2$S\\
1s$^2$2s2p($^3$P$^\mathrm{o}$)${\bar 3}$s & $^4$P$^\mathrm{o}$\\
1s$^2$2s$^2{\bar 3}$p                     & $^2$P$^\mathrm{o}$\\
1s$^2$2s$^2{\bar 3}$d                     & $^2$D\\
1s$^2$2s2p${\bar 3}$p                     & $^2$P\\
1s$^2$2s2p${\bar 3}$s                     & $^2$P$^\mathrm{o}$\\
1s$^2$2s2p${\bar 3}$p                     & $^4$P\\
                                          & $^2$D\\
1s$^2$2s2p${\bar 3}$s                     & $^2$P$^\mathrm{o}$\\
1s$^2$2p$^2$${\bar 3}$p                   & $^4$D$^\mathrm{o}$\\
1s$^2$2s2p${\bar 3}$d                     & $^4$P$^\mathrm{o}$\\
\cline{1-2} \\
\end{tabular}
\end{table}

In Table~\ref{levels}, we compare experimental target state energies
(Eriksson \citealt{Eriksson83}) as well as values calculated by
Kisielius \& Storey \cite{KS02} with our results for the eight lowest
target terms that belong to the three lowest configurations, 2s$^2$2p,
2s2p$^2$ and 2p$^3$. We use the experimental target energies in the
calculation of the Hamiltonian matrix of the $(N+1)$ electron system
and in the calculation of energy levels, oscillator strengths and
photoionization cross-sections of \ion{N}{ii}. We use nineteen target
terms in our R-matrix calculation where we add selected terms from the
seven configurations
1s$^2$2s$^2\,{\bar 3}$s, 1s$^2$2s$^2\,{\bar 3}$p, 1s$^2$2s$^2\,{\bar
  3}$d, 1s$^2$2s2p\,${\bar 3}$s, 1s$^2$2s2p\,${\bar 3}$p,
1s$^2$2s2p\,${\bar 3}$d and 1s$^2$2p$^2$\,${\bar 3}p$ in order to increase 
the dipole polarizability of the terms of the 2s$^2$2p and 2s2p$^2$ 
configurations of the \ion{N}{$^{2+}$} target. These additional terms 
provide the main contributions to the dipole polarizability of the
1s$^2$2s$^2$2p~$^2$P$^{\rm o}$ and 1s$^2$2s2p$^2$~$^4$P states and
significant contributions to the polarizability of
1s$^2$2s2p$^2$~$^2$D, $^2$S and $^2$P. The chosen target terms are
listed in Table~\ref{targets}, which shows that our calculated target
energies are in slightly worse agreement with experiment than those of
Kisielius \& Storey \cite{KS02} although it should be noted that their
energies were calculated in $LS$-coupling, whereas ours are weighted
averages of fine-structure level energies. The configuration
interaction in our target is less extensive than in theirs due to
the additional computational constraints imposed by an intermediate coupling
calculation as opposed to an $LS$-coupling one. We do, however,
consider that our set of target states represents the polarizability of the 
important \ion{N}{$^{2+}$} states better than does the target of Kisielius 
\& Storey \cite{KS02}.

\subsection{\label{s:data2}
Energy levels of \ion{N}{$^+$}}

Experimental energy levels for \ion{N}{$^+$} have been given by Eriksson 
\cite{Eriksson83} for members of the series 2s$^2$2p($^2$P$^{\rm o}$)\,$nl$ 
with $n \leq 16$ and $l \leq 4$, for the series 
2s2p$^2$($^4$P$^{\rm e}$)\,$nl$ with $n \leq 12$ and $l \leq 4$ and for the 
series 2s2p$^2$($^2$D$^{\rm e}$)\,$3l$, $4$p although some levels are missing. 
Energy levels for the states $^3$P$^{\rm e}$$_{2,1,0}$ belonging to the 
equivalent electron configuration 2s$^2$2p$^4$ are also presented.  We use 
these experimental data as a benchmark for our \ion{N}{ii} energy level 
calculation.

The current calculations of energy levels include only the states
belonging to the configurations 2s$^2$2p($^2$P$^{\rm o}_{J_{\rm
    C}}$)\,$nl$, 2s2p$^2$($^4$P$^{\rm e}_{J_{\rm C}}$)\,$nl$ and
possibly 2s2p$^2$($^2$D$^{\rm e}_{J_{\rm C}}$)\,$nl$ with ionization
energies less than $E_0$ and total orbital angular momentum quantum
number $L \leq 8$, where $J_{\rm C}$ is the total angular momentum
quantum number of the core electrons.  Every single energy level
calculated by {\it ab initio} methods is further identified with the
help of the experimental energy level tables of Eriksson
\cite{Eriksson83}, and is used in preference to quantum defect
extrapolation from experimentally known lower states. The bound state
energies for the lowest levels are calculated using R-matrix codes,
with an effective principal quantum number range set to be $0.5 \leq
\nu \leq 10.5$, and the codes are run with a searching step of
$\delta\nu = 0.01$.  We assume that the subsequent iteration converges
on a final energy when $\Delta\,E < 1\times10^{-5}$~Ryd. We only keep
the final energy levels with $n \leq 11$ and $l \leq 4$, and delete
the $n$g levels with $n > 6$, because of numerical instabilities in
the codes.  In total, 377 levels are obtained.  The photoionization
cross-sections from these bound states are calculated later using
outer region R-matrix codes.

For states with $11 < n \leq n_d$, and $l \leq 3$, where calculated
energies exist for lower members of the series, a quantum defect has been
calculated for the highest known member (usually with $n = 11$), and this
quantum defect is used to determine the energies of all higher terms.

Finally, if neither of the above methods can be used, the state is assumed to
have a zero quantum defect.

\subsection{\label{s:data3}
Bound-bound radiative data}

Radiative transition probabilities are taken from three sources,

(1) {\it Ab initio} calculations: We have computed values of weighted
oscillator strengths, $gf$, for all transitions between bound states with 
ionization energies less than or equal to $E_0$ (corresponding to 
$n=11$ in the principal series of \ion{N}{$^{+}$}), with total orbital 
angular momentum quantum number $L \leq 8$, and with total angular momentum 
quantum number $J \leq 6$. The data are calculated in the intermediate coupling
scheme, so there are transitions between states of different total spins. 
Two-electron transitions, which involve a change of core state, are also 
included.

(2) Coulomb approximation: For pairs of levels where oscillator strengths
are not computed by the {\it ab initio} method, but where one or both of the
states have a non-zero quantum defect, the dipole radial integrals required for
the calculation of transition probabilities are calculated using the Coulomb
approximation.  Details are given by Storey \cite{Storey94}.

(3) Hydrogenic approximation: For pairs of levels with a zero quantum defect
hydrogenic dipole radial integrals are calculated, using direct recursion on
the matrix elements themselves as described by Storey \& Hummmer \cite{SH91}.

\subsection{\label{s:data4}
Photoionization cross-sections and recombination coefficients}

The recombination coefficient for each level $nls\,L[K]^{\pi}\,_{J}$ is
calculated directly from the photoionization cross-sections for that state. 
There are three approximations in which the photoionization data are obtained.

(1) Photoionization cross-sections are computed for all the 377 states with
ionization energy less than or equal to $E_0$, $L\,\leq\,8$ and
$J\,\leq\,6$. We obtain the recombination coefficient directly by integrating
the appropriate R-matrix photoionization cross-sections.

(2) Coulomb approximation: As in the bound-bound case, the Coulomb 
approximation is used for states where no OP data are available, but which 
have a non-zero quantum defect.  The calculation of photoionization data using
Coulomb functions has been described by Burgess \& Seaton \cite{BS60} and 
Peach \cite{Peach67}.

(3) For all states for which the R-matrix photoionization cross-sections have 
not been calculated explicitly, the hydrogenic approximation to the 
photoionization cross-sections is evaluated, using the routines of 
Storey \& Hummer \cite{SH91} to generate radiative data in hydrogenic systems.

\subsection{\label{s:data5}
Energy mesh for \ion{N}{ii} photoionization cross-sections}

The photoionization cross-sections for \ion{N}{ii} generated by the Opacity 
Project (OP) method were based on a quantum defect mesh with 100 points per 
unit increase in the effective quantum number derived from the next threshold. 
In contrast to the OP calculations, we use a variable step mesh for 
photoionization cross-section calculations for a particular energy region 
above the $^2$P$^{\rm o}$\,$_{3/2}$ threshold, which is appropriate to 
dielectronic recombination in nebular physical conditions.  This energy 
mesh delineates all resonances to a prescribed accuracy (Kisielius et al. 
\citealt{KS98}; Kisielius \& Storey \citealt{KS02}). 
This detailed consideration of the energy mesh was undertaken for the region 
from the 2s$^2$2p\,($^2$P$^{\rm o}$\,$_{3/2}$) limit up to 0.160 Ryd 
($n = 5$) below the 2s2p$^2$\,($^4$P\,$_{1/2}$) limit, since this region 
contains the main contribution to the total recombination at the temperatures 
of interest for the triplet and singlet series.

From 0.160 Ryd below the 2s2p$^2$\,($^4$P\,$_{1/2}$) limit to 0.0331 Ryd 
below the 2s2p$^2$\,($^4$P\,$_{1/2}$) limit, we use a quantum defect mesh 
with an increment of $0.01$ in effective principal quantum number. The energy 
0.160 Ryd corresponds to a principal quantum number of five relative to the 
next threshold, and 0.0331 Ryd corresponds to eleven.  In total about 600 
points are used in this region. 
For the region from 0.0331 Ryd ($n = 11$) below the 
2s2p$^2$\,($^4$P\,$_{1/2}$) limit up to the 2s2p$^2$\,($^4$P\,$_{5/2}$) limit, 
we use the Gailitis average (Gailitis \citealt{Gailitis}). We use the method 
for this part of photoionization calculations because of the very dense 
resonances in this narrow energy region. About 100 points are used.

In the region from the 2s2p$^2$\,($^4$P\,$_{5/2}$) limit up to 0.0331 Ryd 
($n = 11$) below the 2s2p$^2$\,($^2$D\,$_{3/2}$) limit, a quantum defect 
mesh is again used, with an increment of $0.01$ in effective principal quantum 
number.  Here 0.0331 Ryd corresponds to a principal quantum number of eleven 
relative to the next threshold, 2s2p$^2$\,($^2$D\,$_{3/2}$).  There are about 
800 points used in this region. 
For the energetically lowest region between the two ground fine-structure 
levels of \ion{N}{iii}, 2s$^2$2p\,($^2$P$^{\rm o}$\,$_{1/2}$) and 
2s$^2$2p\,($^2$P$^{\rm o}$\,$_{3/2}$), we use linear extrapolation from 
a few points lying right above the 2s$^2$2p\,($^2$P$^{\rm o}$\,$_{3/2}$) 
threshold.  About 100 points are used in this region.

During the calculation of photoionization cross-sections, we check for every 
bound states to make sure that the cross-section data of different energy areas 
all join smoothly.
Figure~\ref{fig:photoionizationXS_a} shows the photoionization cross-sections 
calculated from the five lowest levels $^3$P$^{\rm e}$\,$_{0}$, 
$^3$P$^{\rm e}$\,$_{1}$, $^3$P$^{\rm e}$\,$_{2}$, $^1$D$^{\rm e}$\,$_{2}$ and 
$^1$S$^{\rm e}$\,$_{0}$ belonging to the ground configuration 
1s$^2$2s$^2$2p$^2$ of \ion{N}{$^+$}. Photoionization cross-sections calculated 
for the five different energy regions have been joined together.

%
\begin{figure}[ht]
\centering
  \includegraphics[width=5.5cm,angle=-90]{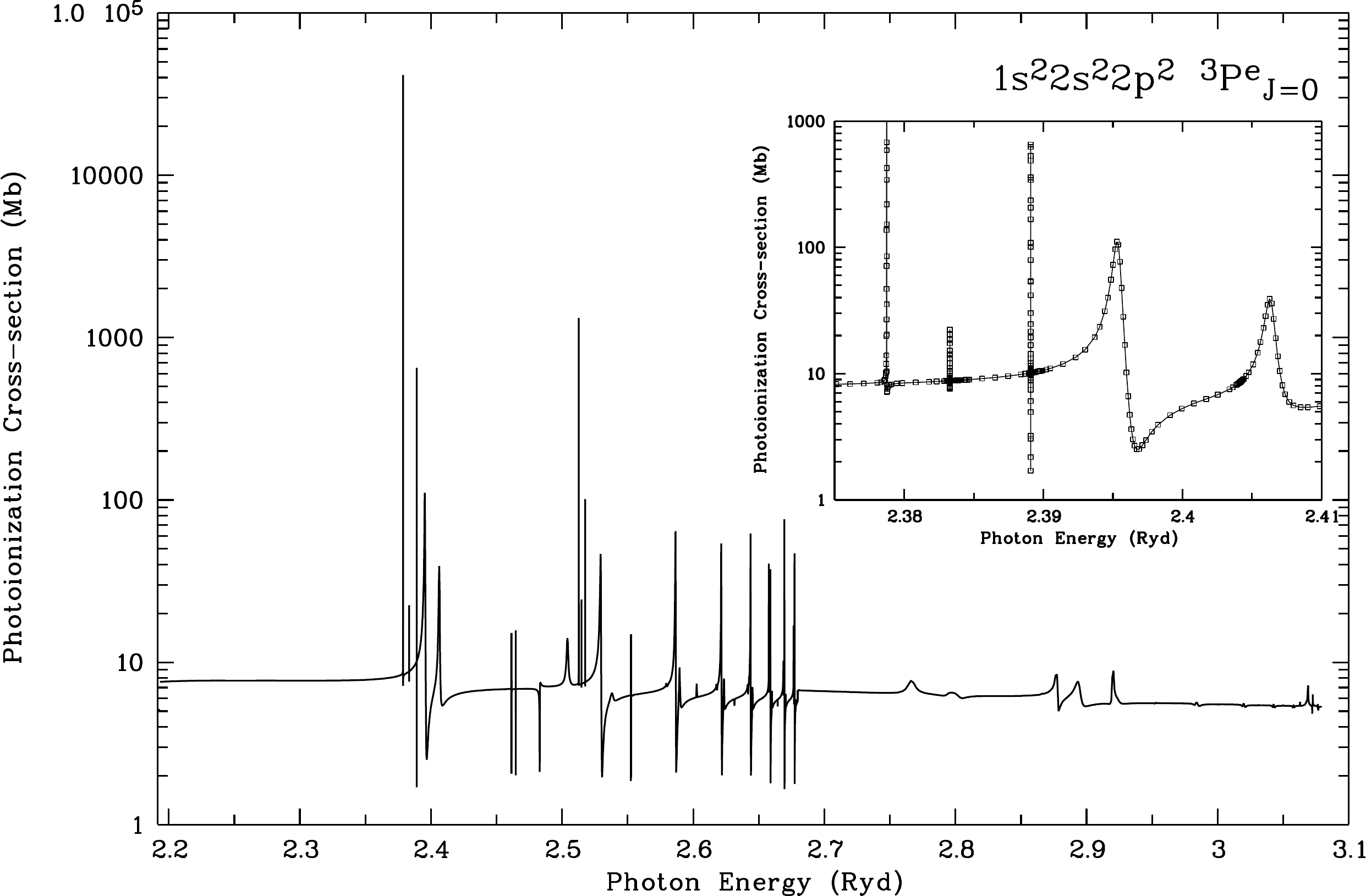}
  \caption{(a) Photoionization cross-sections from the lowest level of
           \ion{N}{ii}: 2s$^2$2p$^2$ $^3$P$^{\rm e}$\,$_{J=0}$. 
           The data calculated by different methods for the five energy 
           regions have been joined together. The insert zooms in a particular 
           energy area, and the mesh points used for the photoionization 
           calculations are shown in the inset.}
  \label{fig:photoionizationXS_a}
\end{figure}
\addtocounter{figure}{-1}
\begin{figure}[ht]
\centering
  \includegraphics[width=5.5cm,angle=-90]{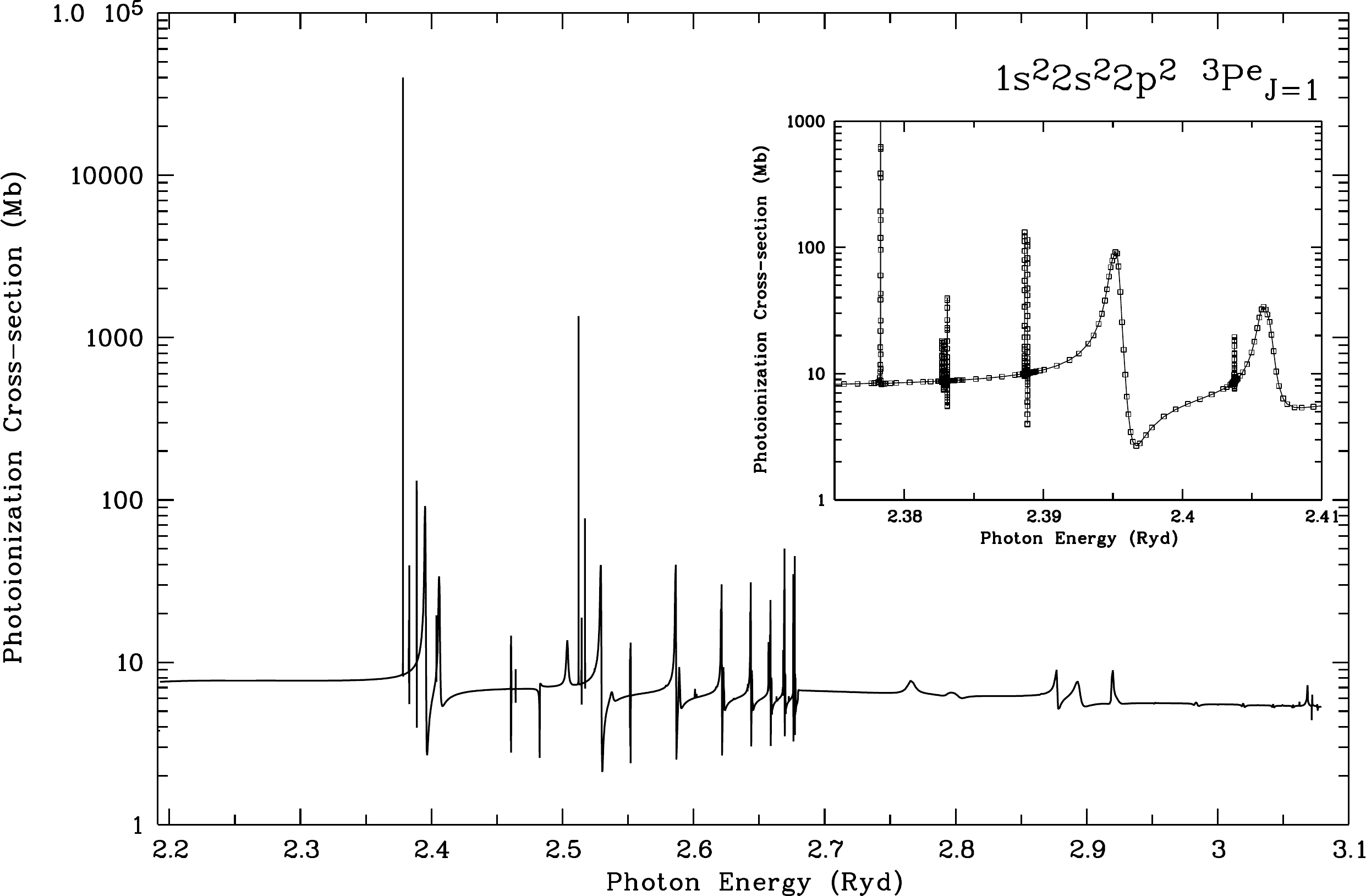}
  \caption{-- {\em Continued.}  (b) Photoionization cross-sections from the 
           level of \ion{N}{ii}: 2s$^2$2p$^2$ $^3$P$^{\rm e}$\,$_{J=1}$. 
           See also caption to Fig.~\ref{fig:photoionizationXS_a} (a).}
  \label{fig:photoionizationXS_b}
\end{figure}
\addtocounter{figure}{-1}
\begin{figure}[ht]
\centering
  \includegraphics[width=5.5cm,angle=-90]{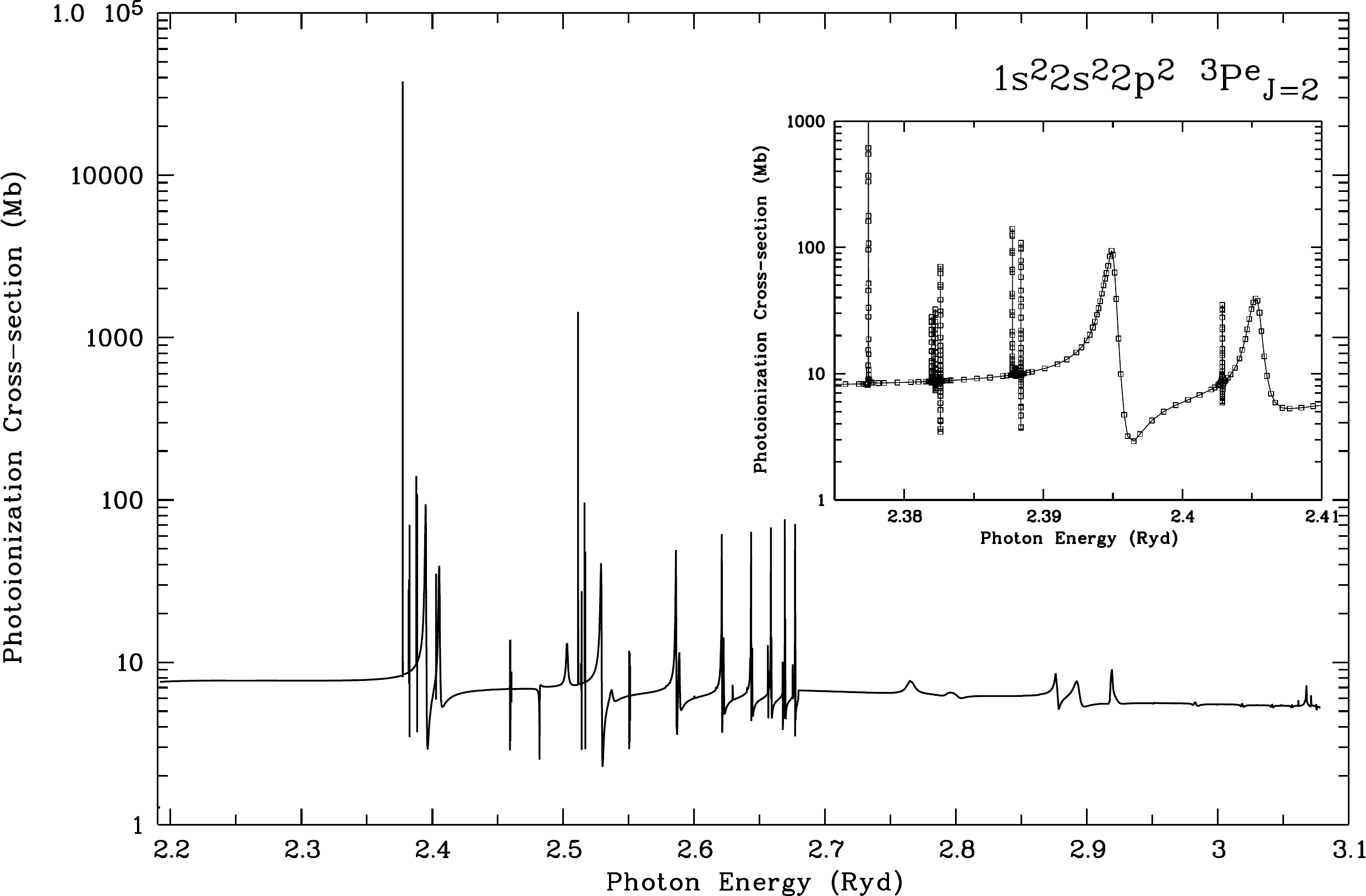}
  \caption{-- {\em Continued.}  (c) Photoionization cross-sections from the 
           level of \ion{N}{ii}: 2s$^2$2p$^2$ $^3$P$^{\rm e}$\,$_{J=2}$. 
           See also caption to Fig.~\ref{fig:photoionizationXS_a} (a).}
  \label{fig:photoionizationXS_c}
\end{figure}
\addtocounter{figure}{-1}
\begin{figure}[ht]
\centering
  \includegraphics[width=5.5cm,angle=-90]{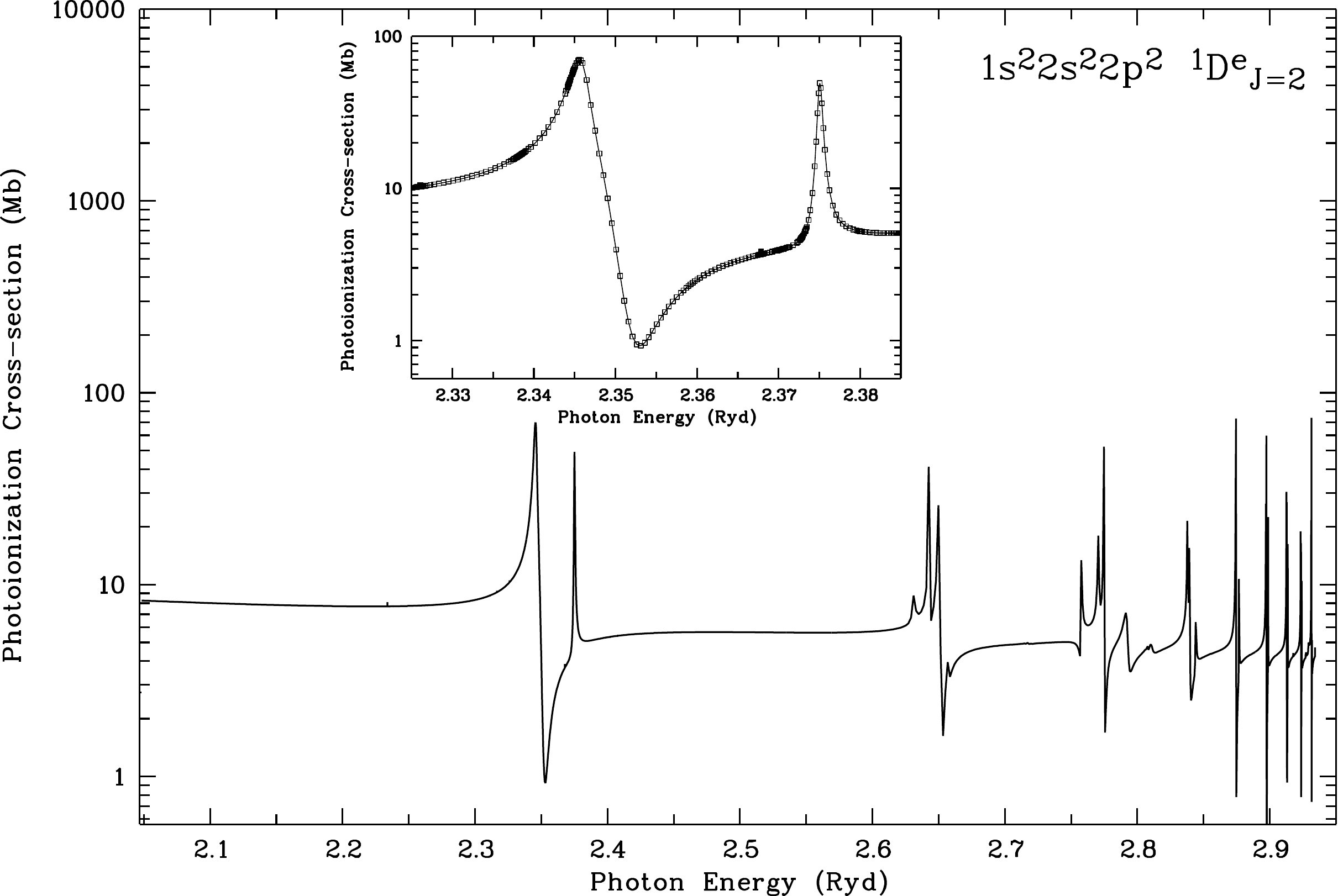}
  \caption{-- {\em Continued.}  (d) Photoionization cross-sections from the 
           level of \ion{N}{ii}: 2s$^2$2p$^2$ $^1$D$^{\rm e}$\,$_{J=2}$. 
           See also caption to Fig.~\ref{fig:photoionizationXS_a} (a).}
  \label{fig:photoionizationXS_d}
\end{figure}
\addtocounter{figure}{-1}
\begin{figure}[ht]
\centering
  \includegraphics[width=5.5cm,angle=-90]{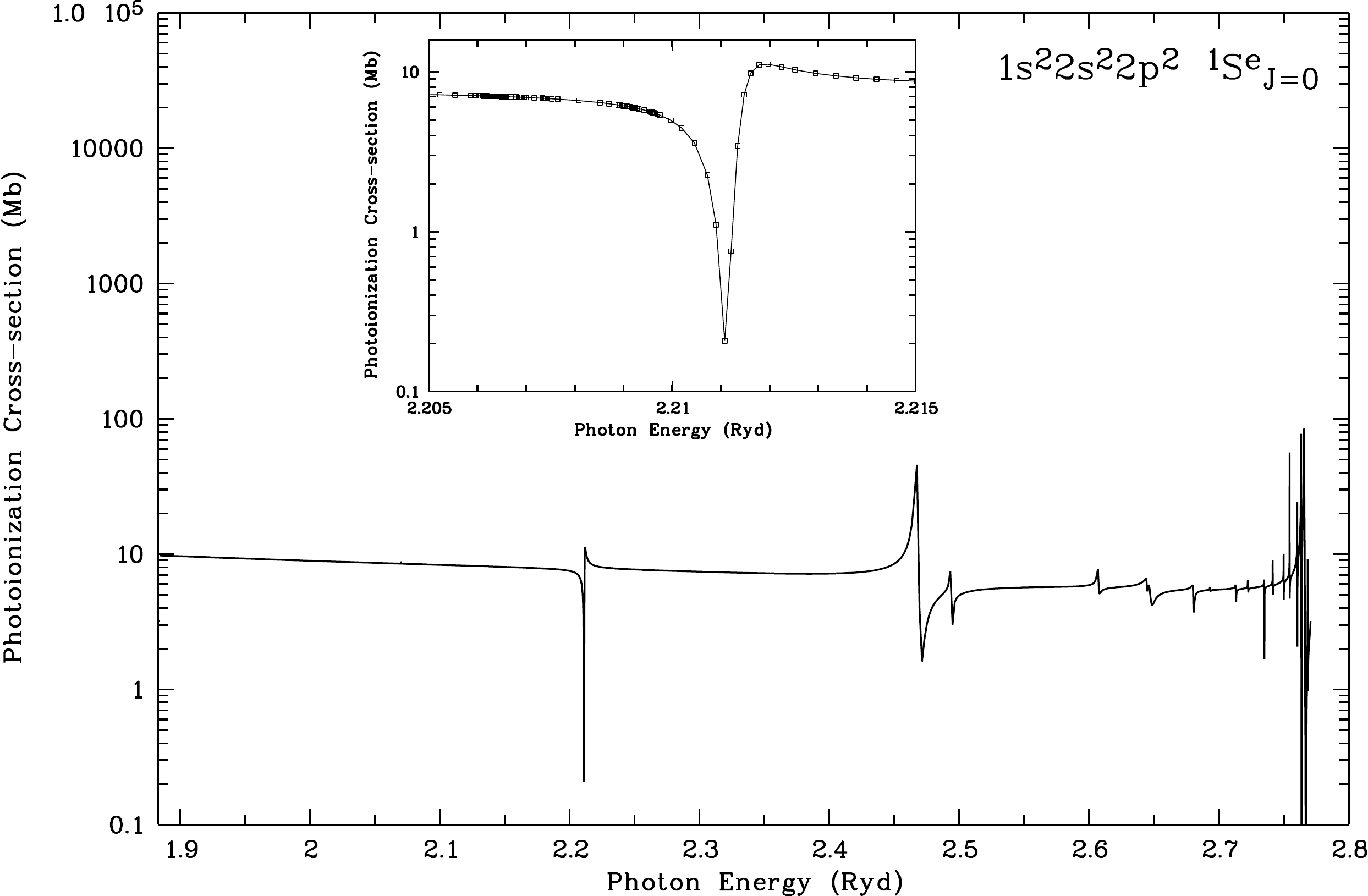}
  \caption{-- {\em Continued.}  (e) Photoionization cross-sections from the 
           level of \ion{N}{ii}: 2s$^2$2p$^2$ $^1$S$^{\rm e}$\,$_{J=0}$. 
           See also caption to Fig.~\ref{fig:photoionizationXS_a} (a).}
  \label{fig:photoionizationXS_e}
\end{figure}

\section{\label{s:population}
Calculation of \ion{N}{$^+$} population}

\subsection{\label{s:population0}
The \ion{N}{$^+$} populations}

The calculation of populations is a three stage process to compute departure 
coefficients, $b(J_{\rm C}; nl\,K\,J)$, defined in terms of populations by 

\begin{equation} \label{eq:popul}
\left( {{N(J_{\rm C}; nl\,K\,J)}\over{N_\mathrm{e} N_+(J_{\rm C})}} \right) =
  \left( {{N(J_{\rm C}; nl\,K\,J)}\over{N_\mathrm{e} N_+(J_{\rm C})}} \right)_S
  \  b(J_{\rm C}; nl\,K\,J),
\end{equation}
where $J_{\rm C}$ is an N$^{2+}$ core state and the subscript $S$ refers to
the value of the ratio given by the Saha and Boltzmann equations, and
$N_\mathrm{e}$ and $N_+(J_{\rm C})$ are the number densities of electrons
and recombining ions, respectively. We distinguish two boundaries in
principal quantum number, $n_d$ and $n_l$. For $n \leq n_d$ collisional
processes are negligible compared to radiative decays and can be
omitted from the calculation of the populations.  For higher $n$ a
full collisional-radiative treatment of the populations is necessary
as described by Hummer \& Storey \cite{HS87} and Storey \& Hummer
\cite{SH95} with some additions to treat dielectronic
recombination. The boundary at $n = n_l$ is defined such that for $n > n_l$ 
the redistribution of population due to $l$-changing collisions
is rapid enough to assume that the populations obey the Boltzmann 
distribution for a given $n$ and hence that $b_{nl} = b_n$ for all $l$. \\

The three stages of the calculation are as follows: \\ 

Stage 1: A calculation of $b(J_{\rm C}; n)$ is made for all $n < 1000$,
using the techniques and atomic rate coefficients described by Hummer
\& Storey \cite{HS87} and Storey \& Hummer \cite{SH95} with the
addition of $l$-averaged autoionization and dielectronic capture rates
computed with AUTOSTRUCTURE for states of ($^2$P$^{\rm o}$\,$_{3/2}$)
parentage that lie above the ionization limit. For $n > 1000$, we
assume $b_n=1$. The results of this calculation provide the values of
$b$ for $n > n_l$ and the initial values for $n \leq n_l$ for Stage 2. \\

Stage 2: A calculation of $b(J_{\rm C}; nl\,K\,J)$ is made for all $n \leq n_l$
using the same collisional-radiative treatment as in Stage 1 but now
resolved by total $J$. The combined results of Stages 1 and 2 provide
the values of $b$ for $n > n_d$. \\

Stage 3: For the energies less than that which corresponds to $n = n_d$ in 
the principal series, departure coefficients $b(J_{\rm C}; nl\,K\,J)$ are 
computed for states of all series. Since only spontaneous radiative decays 
link these states, the populations are obtained by a step-wise solution from 
the energetically highest to the lowest state.

\subsection{\label{s:population1}
Dielectronic recombination within the $^2$P$^{\mathrm o}$ parents}

\begin{figure*}[ht]
\centering
  \includegraphics[width=14cm,angle=0]{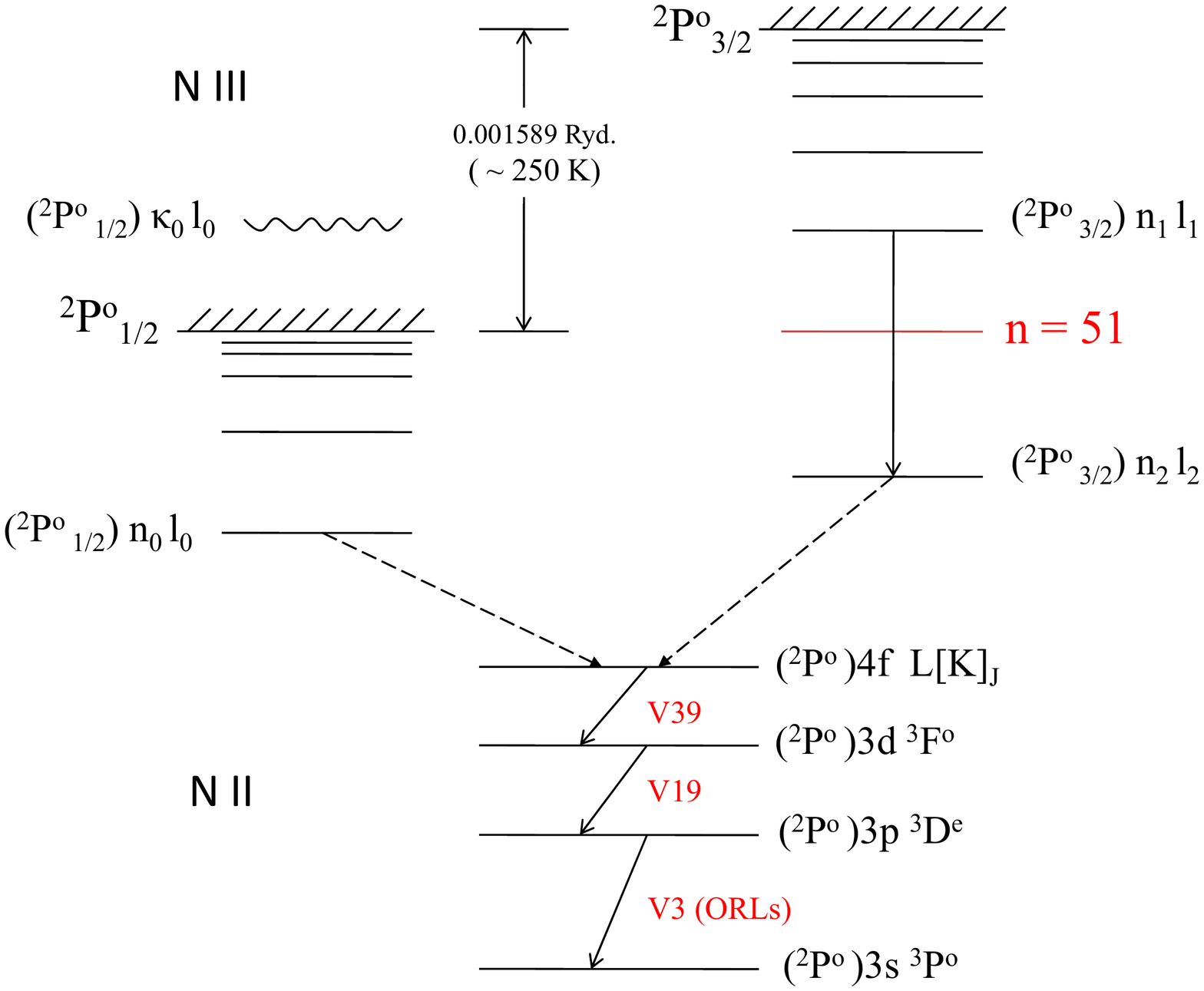}
  \caption{Schematic figure showing the low-temperature ($\leq 250$~K) 
           dielectronic recombination of \ion{N}{ii} through the fine-structure 
           autoionizing levels between the two lowest ionization 
           thresholds of \ion{N}{iii} $^2$P$^{\rm o}$\,$_{1/2}$ and 
           $^2$P$^{\rm o}$\,$_{3/2}$.  The electrons captured to the 
           high-$n$ autoionizing levels decay to low-$n$ bound states through 
           cascades and optical recombination lines (ORLs) are thus emitted.
           Here Multiplets V3, V19 and V39 are presented as examples.}
  \label{fig:2PoScheme}
\end{figure*}

Within the $^2$P$^{\rm o}$ parents, the contribution by dielectronic
recombination to the total recombination is shown to become significant at 
very low temperatures ($\leq 250$~K), due to recombination into high-lying 
bound states of the $^2$P$^{\rm o}_{3/2}$ parent from the $^2$P$^{\rm o}_{1/2}$
continuum states. We incorporate this low-temperature process into our 
calculation of the \ion{N}{$^+$} populations. Figure~\ref{fig:2PoScheme} is a 
schematic diagram illustrating \ion{N}{ii} dielectronic capture, 
autoionization and radiative decays. The electrons captured to the 
high-n autoionizing levels decay to low-n bound states through cascades, and 
optical recombination lines are emitted. Radiative transitions which change
parent, such as ($^2$P$^{\rm o}_{3/2}$)$n_1l_1$ -- ($^2$P$^{\rm
 o}_{1/2}$)$n_0l_0$, are included for those states for which R-matrix
calculated values are present. Higher states are treated by hydrogenic or 
Coulomb approximations which do not allow parent changing transitions.

In Figure~\ref{fig:2PoScheme}, three multiplets of \ion{N}{ii} are presented 
as examples: V3 2s$^2$2p3p\,$^3$D$^{\rm e}$ -- 2s$^2$2p3s\,$^3$P$^{\rm o}$, 
the strongest 3p -- 3s transition, V19 2s$^2$2p3d\,$^3$F$^{\rm o}$ -- 
2s$^2$2p3p\,$^3$D$^{\rm e}$, the strongest 3d -- 3p transition, and V39 
2s$^2$2p4f\,G[7/2,9/2]$^{\rm e}$ -- 2s$^2$2p3d\,$^3$F$^{\rm o}$, the strongest
4f -- 3d transition. The results for these three multiplets are analysed in 
Section~\ref{s:result2}.

\subsection{\label{s:population2}
The $^2$P$^{\rm o}$ parent populations}

The contribution to the total recombination coefficient of a state
depends on the relative populations of the $^2$P$^{\rm o}$\,$_{1/2}$
and $^2$P$^{\rm o}$\,$_{3/2}$ parent levels, which generally dominate
the populations of the recombining ion \ion{N}{$^{2+}$} under typical
nebular physical conditions. The relative populations of the 
two fine-structure levels deviate from the statistical weight ratio, 
$1\,:\,2$, which is assumed in all work hitherto on this ion. 
The deviation affects the populations of the high Rydberg states, and 
consequently total dielectronic recombination coefficients at low-density 
and low-temperature conditions.

We model the \ion{N}{$^{2+}$} populations with a five level atom
comprising the two levels of the $^2$P$^{\rm o}$ term and the three
levels of the $^4$P term, although it should be noted that the
populations of the three $^4$P levels are almost negligible in the
nebular conditions considered here (Section~\ref{s:result4}). The 
relative populations are assumed to be determined only by collisional 
excitation, collisional de-excitation and spontaneous radiative decay. 
Transition probabilities were taken from Fang et al. \cite{fang93} 
and thermally averaged collision strengths from Nussbaumer \& Storey 
\cite{NS79} and Butler \& Storey (private communication).

\begin{figure*}[ht]
\centering
  \includegraphics[width=12cm,angle=-90]{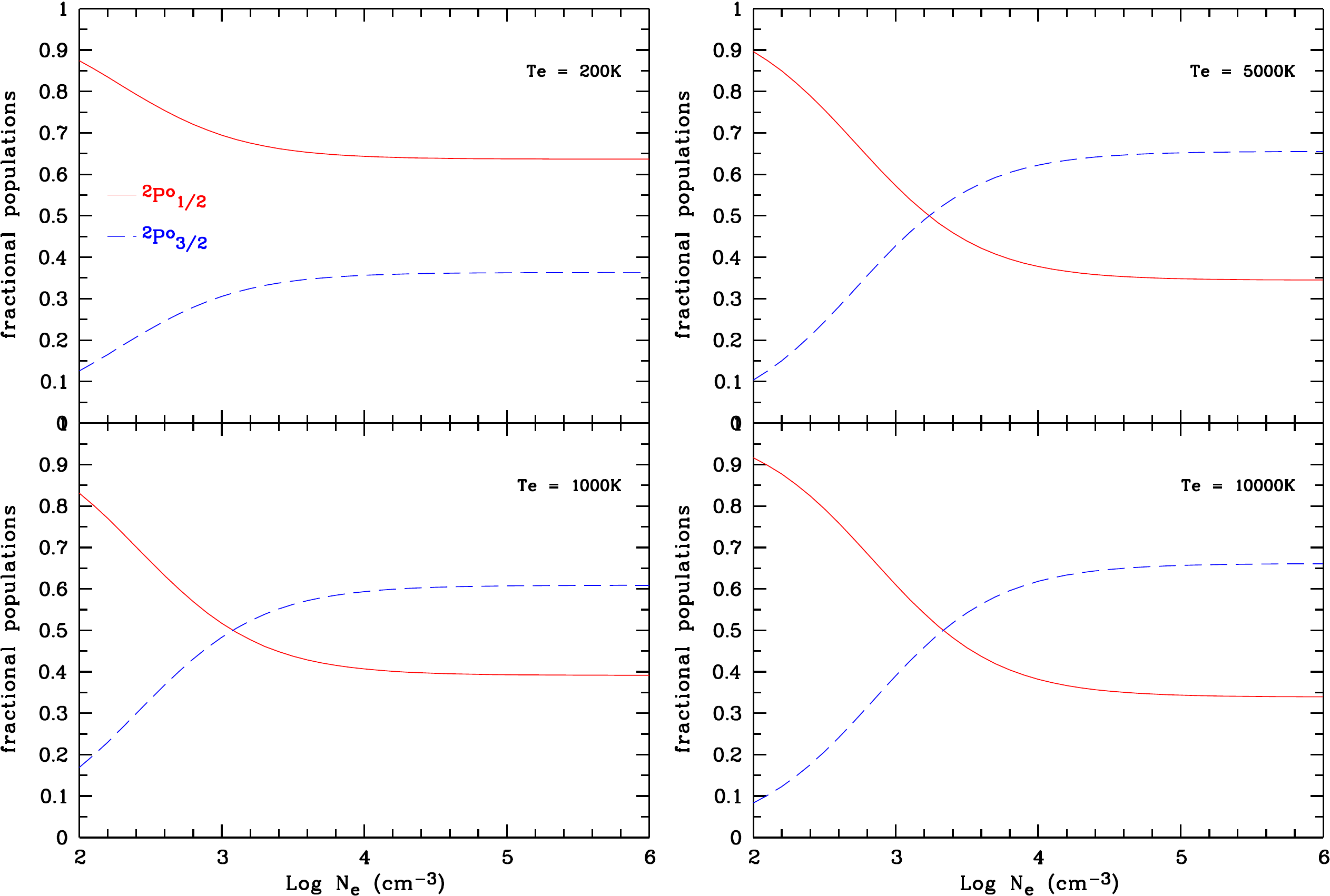}
  \caption{Fractional populations of the \ion{N}{$^{2+}$} 
           $^2$P$^{\rm o}\,_{1/2}$ and $^2$P$^{\rm o}\,_{3/2}$ parent levels. 
           The red solid and blue dashed curves represent the populations 
           of the $^2$P$^{\rm o}$\,$_{1/2}$ and $^2$P$^{\rm o}$\,$_{3/2}$ 
           levels, respectively. Four temperature cases, $200$~K, $1,000$~K, 
           $5,000$~K and $10,000$~K are shown.}
  \label{fig:fractional_population}
\end{figure*}

Figure~\ref{fig:fractional_population} shows the fractional
populations of the \ion{N}{$^{2+}$} $^2$P$^{\rm o}\,_{1/2}$ and
$^2$P$^{\rm o}\,_{3/2}$ fine-structure levels at several electron
temperatures and as a function of electron density, ranging from
10$^2$ to 10$^6$~cm$^{-3}$, applicable to PNe and \ion{H}{ii}
regions. The fractional populations vary significantly below
10$^{4}$~cm$^{-3}$ and converge to the thermalized values at higher
densities.

\subsection{\label{s:population3}
The Cases A and B}

Baker \& Menzel \cite{BM38} define the Cases~A and B with reference
to the recombination spectrum of hydrogen.  In \ion{N}{ii}, there are five 
low-lying levels belonging to the ground configuration 2s$^2$2p$^2$, 
$^3$P$^{\rm e}$\,$_{0,1,2}$, $^1$D$^{\rm e}$\,$_{2}$ and 
$^1$S$^{\rm e}$\,$_{0}$.  Just as in Kisielius \& Storey \cite{KS02}, we 
define two cases for \ion{N}{ii}. In Case~A, all emission lines are assumed 
to be optically thin. In Case~B, lines terminating on the three lowest levels
$^3$P$^{\rm e}$\,$_{0,1,2}$ are assumed to be optically thick and no radiative
decays to these levels are permitted when calculating the population 
structure. The latter case is generally a better approximation for most 
nebulae.

\section{\label{s:results}
Results and discussion}

\subsection{\label{s:result0}
Effective recombination coefficients}

The population structure of \ion{N}{$^+$} has been calculated for electron 
temperature $\log{T_\mathrm{e}[\rm K]} = 2.1 \sim 4.3$, with a step of $0.1$ 
in logarithm, and for the electron density 
$N_\mathrm{e} = 10^2 \sim 10^6$~cm$^{-3}$, also with a step of $0.1$ in 
logarithm. Constrained by the range of the exponential factors 
involved in the calculation of the departure coefficients using the 
Saha-Boltzmann equation, calculation of the effective recombination 
coefficients starts from 
$125$~K ($\log{T_\mathrm{e}}$ = 2.1). For electron densities greater than 
$10^6$~cm$^{-3}$, as pointed out by Kisielius \& Storey \cite{KS02}, it is 
necessary to include $l$-changing collisions for $n < 11$. This is however 
beyond the scope of the current treatment.

In Tables~\ref{aeff_Ne100}, \ref{aeff_Ne1000}, \ref{aeff_Ne10000} and 
\ref{aeff_Ne100000} we present the effective recombination coefficient, 
$\alpha_\mathrm{eff}(\lambda)$, in units of cm$^3$~s$^{-1}$, for strongest 
\ion{N}{ii} transitions with valence electron orbital angular momentum 
quantum number $l \leq 5$, at electron densities $N_\mathrm{e}$=10$^2$, 
10$^3$, 10$^4$ and 10$^5$ cm$^{-3}$, respectively, in Case~B. The effective 
recombination coefficient is defined such that the emissivity 
$\epsilon(\lambda)$, in a transition of wavelength $\lambda$ is given by 

\begin{equation}
  \label{emissivity}
\epsilon(\lambda) = N_\mathrm{e} N_+ \alpha_\mathrm{eff}(\lambda)
{{hc}\over{\lambda}} \; \; \; \; [\rm{ergs\, cm}^{-3} {\rm s}^{-1}].
\end{equation}

Transitions included in the tables are selected according to the following 
criteria:

(1) $\lambda \geq~912$ \AA; 

(2) $\alpha_\mathrm{eff}(\lambda) \geq 1.0 \times 10^{-14}$ cm$^{3}$~s$^{-1}$ 
at $T_\mathrm{e}$ = 1000~K for all $N_\mathrm{e}$'s, and 
$\geq 1.0 \times 10^{-15}$ cm$^{3}$~s$^{-1}$ at all $T_\mathrm{e}$'s and 
$N_\mathrm{e}$'s; 

(3) All fine-structure components are presented for multiplets from the 
3d - 3p and 3p - 3s configurations. For the 4f - 3d configurations, a few 
selected multiplets are listed, but only V38 and V39 includes all the 
individual components. These transitions fall in the visible part of the 
spectrum and among them are the strongest recombination lines of \ion{N}{ii}.

In Tables~\ref{aeff_Ne100}, \ref{aeff_Ne1000}, \ref{aeff_Ne10000} and 
\ref{aeff_Ne100000}, the wavelengths of all the 4 - 3 and 3 - 3 transitions 
and majority of the 5 - 4 and 5 - 3 transitions are experimentally known. 
All the wavelengths of the 6 - 5 and 6 - 4 transitions are predicted. Our 
calculated wavelengths, derived from the experimental energies, agree with 
the experimentally known wavelengths within $0.001\%$. Our predicted 
wavelengths for experimentally unknown transitions agree with those predicted 
by Hirata \& Horaguchi \cite{hh1995} within $0.007\%$ except for one 6d - 3p 
transition wavelength which differs by 0.24\,{\AA}. However, this transition 
is spectroscopically less important compared to the 4 - 3 and 3 - 3 ones.

In these tables, we use the pair-coupling notation $L[K]^{\pi}_{J}$ for the 
states belonging to the ($^2$P$^{\rm o}$)\,$n$f and $n$g configurations, as 
in Eriksson \cite{Eriksson83}. As shown by Cowan \cite{Cowan81}, pair-coupling
is probably appropriate for the states of intermediate-$l$ ($l = 3, 4$). The 
same notation is adopted for the ($^2$P$^{\rm o}$)\,$n$h configurations. For 
states belonging to low-$l$ ($l \leq 2$) configurations, $LS$-coupling 
notation $^{2S+1}L^{\pi}_{J}$ is used.

\addtocounter{table}{1}

\addtocounter{table}{1}

\addtocounter{table}{1}

\addtocounter{table}{1}

\subsection{\label{s:result1}
Effective recombination coefficient fits}

We fit the effective recombination coefficients as a function of electron 
temperature in logarithmic space with analytical expressions for selected 
transitions, using a non-linear least-square algorithm. 
Tables~\ref{fit1_lowTe}, \ref{fit1_highTe}, \ref{fit2_lowTe}, 
\ref{fit2_highTe}, \ref{fit3_lowTe}, \ref{fit3_highTe}, \ref{fit4_lowTe} and 
\ref{fit4_highTe} present fit parameters and maximum deviations $\delta$[\%] 
for four densities, $N_\mathrm{e}$ = 10$^2$, 10$^3$, 10$^4$ and 10$^5$ 
cm$^{-3}$. Only strongest optical transitions are presented, including 
multiplets V3, V5, V19, V20, V28, V29, V38 and V39. As the dependence of 
recombination coefficient on $T_\mathrm{e}$ at electron temperatures below 
$10,000$~K is different from that at high temperatures 
($10,000 \sim 20,000$~K in our case), we use different expressions for the 
two temperature regimes.

For the low-temperature regime, $T_\mathrm{e} < 10,000$~K, effective 
recombination coefficients are dominated by contribution from radiative 
recombination $\alpha_{\rm rad}$, which has a relatively simple dependence 
on electron temperature, $\alpha_{\rm rad} \propto T_\mathrm{e}$$^{-a}$, 
where $a \sim 1$. At low temperatures, dielectronic recombination through 
low-lying autoionizing states are also important for ions such as \ion{C}{ii},
\ion{N}{ii}, \ion{O}{ii}, \ion{Ne}{ii} (Storey \citealt{Storey81}, 
\citealt{Storey83}; Nussbaumer \& Storey \citealt{NS83}, \citealt{NS84}, 
\citealt{NS86}, \citealt{NS87}). Considering the fact that direct radiative 
recombination rate is nearly a linear function of temperature in 
logarithm, and the deviation introduced by dielectronic recombination, we 
use a five-order polynomial expression to fit the effective recombination 
coefficient, 

\begin{equation}
\label{eq:fitt1}
\alpha = a_0 + a_1\,t + a_2\,t^2 + a_3\,t^3 + a_4\,t^4 + a_5\,t^5,
\end{equation}
where $\alpha = \log_\mathrm{10}\,\alpha_\mathrm{eff} + 15$ and 
$t = \log_\mathrm{10}\,T_\mathrm{e}$, and $a_0$, $a_1$, $a_2$, $a_3$, $a_4$ 
and $a_5$ are constants.

For the high-temperature regime, $10,000 \leq T_\mathrm{e} \leq 20,000$~K, 
the contribution from dielectronic recombination, $\alpha_{\rm DR}$,
can significantly exceed that of direct radiative recombination (Burgess 
\citealt{Burgess64}). Dielectronic recombination coefficient 
$\alpha_{\rm DR}$ has a complex exponential dependence on 
$T_\mathrm{e}$ (Seaton \& Storey \citealt{SS76}; Storey \citealt{Storey81}),
$\alpha_{\rm DR}$ $\propto$ $T_\mathrm{e}^{-3/2}$\,exp(-$E/k\,T_\mathrm{e}$), 
where $E$ is the excitation energy of an autoionizing state, to which a free 
electron is captured, relative to the ground state of the recombining ion 
(\ion{N}{$^{2+}$} in our case) and $k$ is the Boltzmann constant. The 
expression adopted for this temperature regime is, 

\begin{equation}
\label{eq:fitt2}
\alpha = (b_0 + b_1\,t + b_2\,t^2 + b_3\,t^3 + b_4\,t^4) \times t^{b_5} \times \exp(b_6\,t),
\end{equation}
where $\alpha = \log_\mathrm{10}\,\alpha_\mathrm{eff} + 15$ and 
$t = T_\mathrm{e}[\rm K] /10^4$, the reduced electron temperature, and 
$b_0$, $b_1$, $b_2$, $b_3$, $b_4$, $b_5$ and $b_6$ are constants.

\begin{figure*}[ht]
\centering
  \includegraphics[width=7cm,angle=-90]{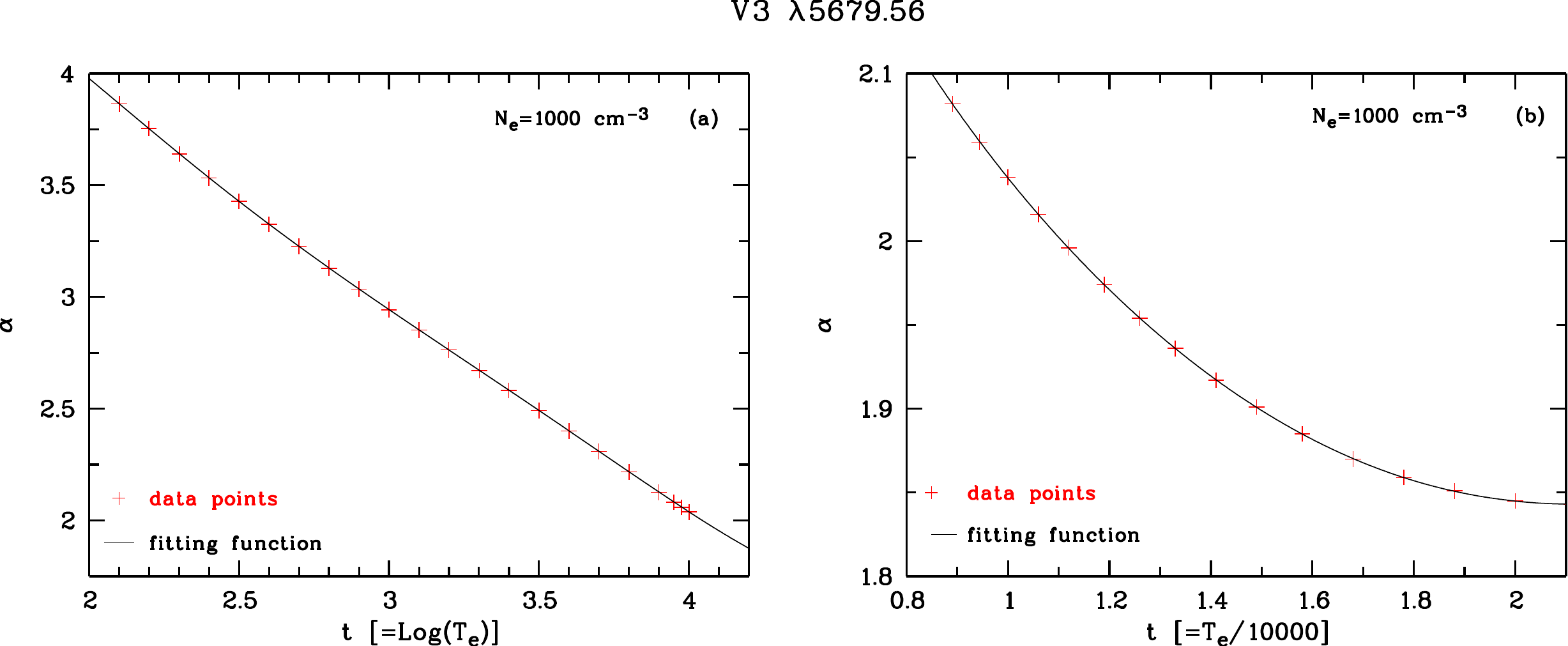}
  \caption{Analysis fit to the effective recombination coefficients for the 
           \ion{N}{ii} V3 3p~$^3$D$_{3}$ -- 3s~$^3$P$^{\rm o}_{2}$ 
           $\lambda$5679.56 line, at $N_\mathrm{e}$ = 1000~cm$^{-3}$. 
           Sub-figure (a) shows the data fit for the low-temperature regime 
           $T_\mathrm{e} < 10,000$~K, and Sub-figure (b) for the 
           high-temperature regime $10,000 \leq T_\mathrm{e} \leq 20,000$~K. 
           In both Sub-figures, solid lines are fitting equations (Equ.~8 for
           Sub-figure (a) and Equ.~9 for (b)), and plus signs ``+" are the 
           calculated data for all the temperatures.}
  \label{fig:data_fits}
\end{figure*}

In order to make the data fits accurate for the high-temperature
regime, $10,000 \leq T_\mathrm{e} \leq 20,000$~K, where the original
calculations are carried out for only four temperature cases
($\log{T_\mathrm{e}[\rm K]}$ = 4.0, 4.1, 4.2 and 4.3), nine more
temperature cases are calculated. For the temperature region
$\log{T_\mathrm{e}[\rm K]}$ = 3.9$\sim$4.0, two more temperature cases
are also calculated, so that the data fits near 10,000~K are accurate
enough. Figure\,\ref{fig:data_fits} is an example of the fit to the
effective recombination coefficients of the \ion{N}{ii} V3
2s$^2$2p3p\,$^3$D$^{\rm e}_{3}$ -- 2s$^2$2p3s\,$^3$P$^{\rm o}_{2}$
$\lambda$5679.56 transition.

By using different expressions for the two temperature regimes, we manage 
to control the maximum fitting errors to well within 0.5 per cent.

\addtocounter{table}{1}

\addtocounter{table}{1}

\addtocounter{table}{1}

\addtocounter{table}{1}

\addtocounter{table}{1}

\addtocounter{table}{1}

\addtocounter{table}{1}

\addtocounter{table}{1}

\subsection{\label{s:result2}
Relative intensities within \ion{N}{ii} multiplets}

As mentioned in the Section~\ref{s:population3} above, the populations of the 
ground fine-structure levels $^2$P$^{\rm o}$\,$_{1/2, 3/2}$ of the recombining 
ion \ion{N}{$^{2+}$} vary with electron density under typical nebular 
conditions.  The variations are reflected in the relative intensities of the 
resultant recombination lines of \ion{N}{ii}, which arise from upper levels 
with the same orbital angular momentum quantum number $l$ but of different 
parentage, i.e., $^2$P$^{\rm o}$\,$_{1/2}$ and $^2$P$^{\rm o}$\,$_{3/2}$ in 
the current case. A number of such recombination lines have been observed in 
photoionized gaseous nebulae including PNe and \ion{H}{ii} regions, and their 
intensity ratios can thus be used for density diagnostics.

As the relative populations of $^2$P$^{\rm o}$\,$_{1/2, 3/2}$ vary with 
$N_\mathrm{e}$, so do the fractional intensities of individual 
fine-structure components within a given multiplet of \ion{N}{ii}. The most 
prominent \ion{N}{ii} multiplets in the optical include: 
V3 2s$^2$2p3p\,$^3$D$^{\rm e}$ -- 2s$^2$2p3s\,$^3$P$^{\rm o}$, 
V19 2s$^2$2p3d\,$^3$F$^{\rm o}$ -- 2s$^2$2p3p\,$^3$D$^{\rm e}$ and 
V39 2s$^2$2p4f\,G[7/2,9/2]$^{\rm e}$ -- 2s$^2$2p3d\,$^3$F$^{\rm o}$.

\subsubsection{\label{V3}
2s$^2$2p3p\,$^3$D$^{\rm e}$ -- 2s$^2$2p3s\,$^3$P$^{\rm o}$ (V3)}

\begin{figure*}[ht]
\centering
   \includegraphics[width=12cm,angle=-90]{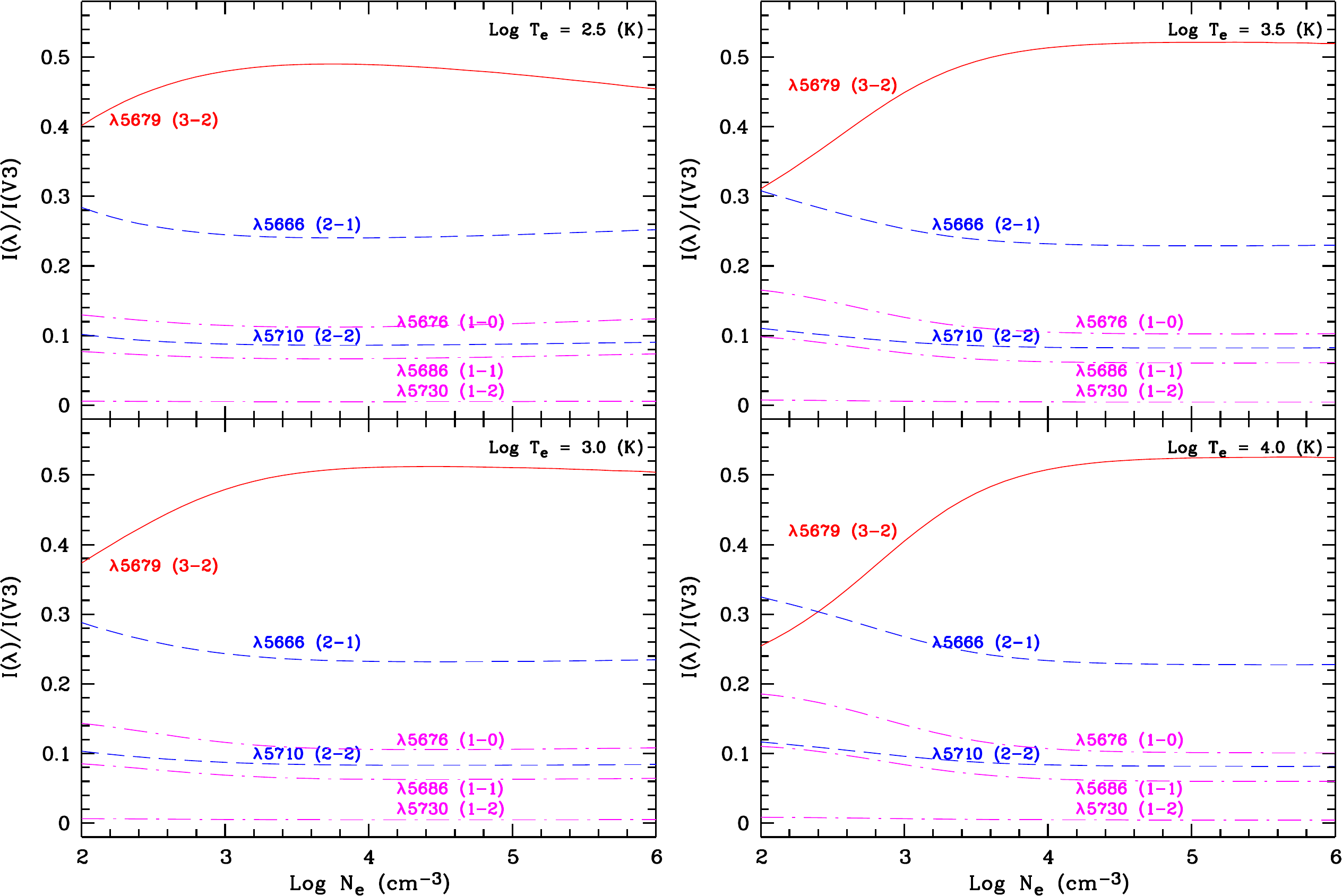}
   \caption{Fractional intensities of components of Multiplet V3: 
            2s$^2$2p3p\,$^3$D$^{\rm e}$ -- 2s$^2$2p3s\,$^3$P$^{\rm o}$. The 
            numbers in brackets following the wavelength labels are the 
            total angular momentum quantum numbers $J_2 - J_1$ of the upper 
            to lower levels of the transition. Components from upper levels 
            of the same total angular momentum quantum number $J$ are 
            represented by same colour and line type. Four temperature cases,
            $\log$$_{10}$T$_{\rm e}$ = 2.5, 3.0, 3.5 and 4.0~K, are 
            presented.}
   \label{multipletV3}
\end{figure*}

The fractional intensities of fine-structure components of Multiplet V3, 
2s$^2$2p3p\,$^3$D$^{\rm e}$ -- 2s$^2$2p3s\,$^3$P$^{\rm o}$, are presented 
in Figure~\ref{multipletV3}. The strongest component is $\lambda$5679.56, 
which forms from core $^2$P$^{\rm o}$\,$_{3/2}$ capturing an electron 
plus cascades from higher states, while the second strongest component 
$\lambda$5666.63 can form, in addition, from recombination of core 
$^2$P$^{\rm o}$\,$_{1/2}$.

For the target \ion{N}{iii}, the population of the fine-structure
level $^2$P$^{\rm o}$\,$_{3/2}$ relative to $^2$P$^{\rm o}$\,$_{1/2}$
increases with electron density $N_\mathrm{e}$ due to collisional
excitation, and consequently, so does the intensity of the
$\lambda$5679.56 line relative to the $\lambda$5666.63 line. Their
intensity ratio peaks around $N_\mathrm{e} = 2,000$~cm$^{-3}$, the
critical density $N_{c}$ of the level $^2$P$^{\rm o}$\,$_{3/2}$, and
then decreases as $N_\mathrm{e}$ increases further. The relative
intensities of all components converge to constant values at high
densities ($\geq 10^{5}$ -- $10^{6}$~cm$^{-3}$), as the relative
populations of the ground fine-structure levels of the target
\ion{N}{iii} approach the Boltzmann distribution.

The line ratio $I(\lambda5679.56)$/$I(\lambda5666.63)$ thus serves as 
a density diagnostic for nebulae of low and intermediate densities, 
$N_\mathrm{e} \leq 10^{5}$~cm$^{-3}$. At very low electron
temperatures, where $kT_\mathrm{e}$ is comparable to the $^2$P$^{\rm
  o}$\,$_{1/2}$ - $^2$P$^{\rm o}$\,$_{3/2}$ energy separation, the
sensitivity to density in the components of V3 is reduced. This arises
because, at very low temperatures, the states ($^2$P$^{\rm
  o}$\,$_{3/2}$)\,nl are populated more significantly by dielectronic
capture from the ($^2$P$^{\rm o}$\,$_{1/2}$)\,$\kappa l$ continuum
than by direct recombination on N$^{2+}$\,($^2$P$^{\rm
  o}$\,$_{3/2}$). The density dependence of the population
distribution between the $^2$P$^{\rm o}$\,$_{1/2}$ and $^2$P$^{\rm
  o}$\,$_{3/2}$ is then of less importance. 

\subsubsection{\label{V19}
2s$^2$2p3d\,$^3$F$^{\rm o}$ -- 2s$^2$2p3p\,$^3$D$^{\rm e}$ (V19)}

\begin{figure*}[ht]
\centering
   \includegraphics[width=12cm,angle=-90]{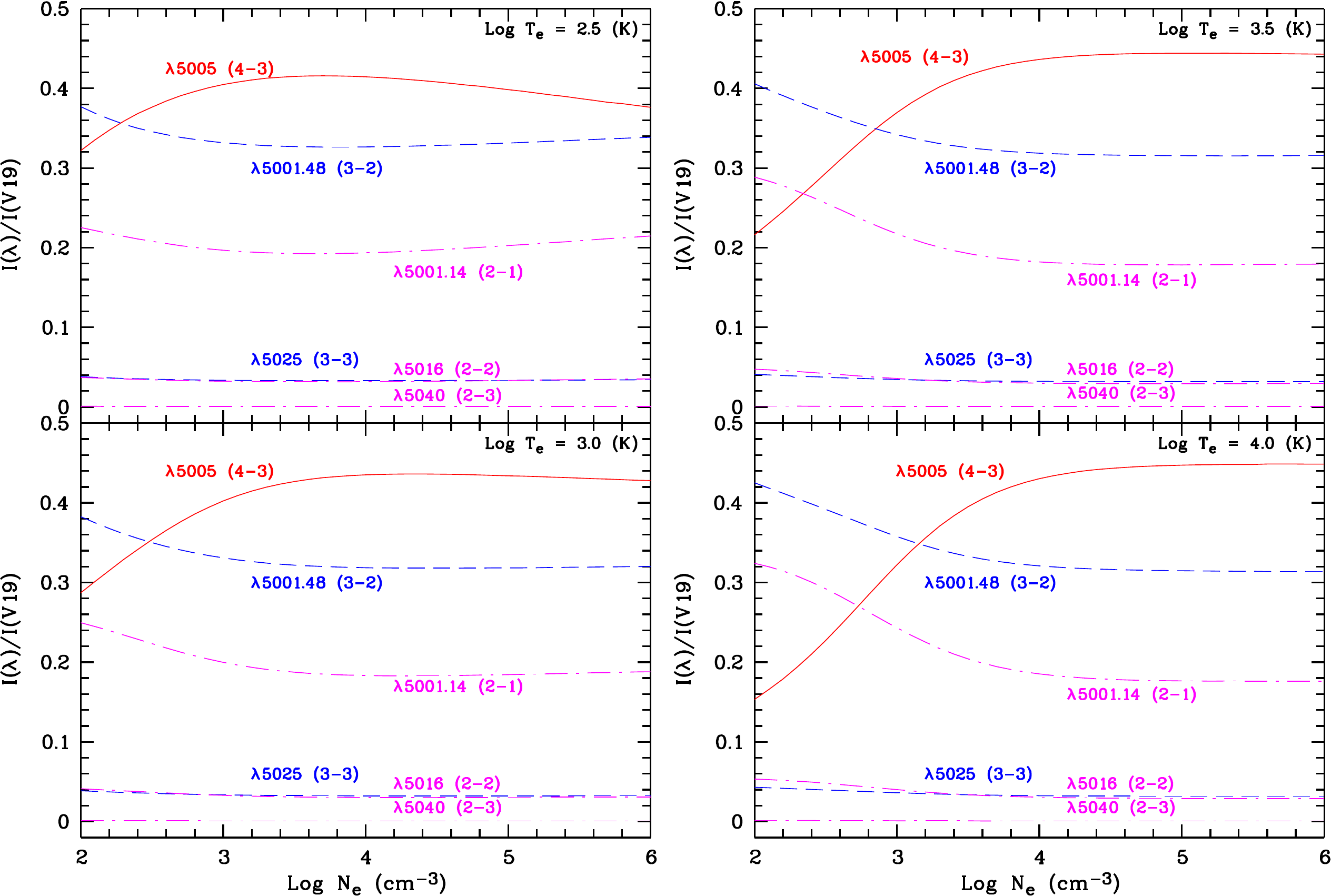}
   \caption{Same as Fig.~\ref{multipletV3} but for Multiplet V19:
            2s$^2$2p3d\,$^3$F$^{\rm o}$ -- 2s$^2$2p3p\,$^3$D$^{\rm e}$.}
   \label{multipletV19}
\end{figure*}

The fractional intensities of fine-structure components of Multiplet V19, 
2s$^2$2p3d\,$^3$F$^{\rm o}$ -- 2s$^2$2p3p\,$^3$D$^{\rm e}$, are presented 
in Figure~\ref{multipletV19}. The strongest component, $\lambda$5005.15, 
forms exclusively from recombination of target $^2$P$^{\rm o}$\,$_{3/2}$ 
plus cascades, while the second and third strongest components of almost 
identical wavelengths, $\lambda$5001.48 and $\lambda$5001.14, can form, in 
addition, from recombination of the ground target $^2$P$^{\rm o}$\,$_{1/2}$.

At very low densities of about $10^2$~cm$^{-3}$, $^2$P$^{\rm o}$\,$_{1/2}$ 
dominates the population of \ion{N}{iii}, and this is manifested by the 
intensity of the $\lambda$5005.15 line being lower than the $\lambda$5001.48
line and than $\lambda$5001.14 by a further amount. As electron 
density increases, the intensity of the $\lambda$5005.15 line relative to the 
$\lambda$5005.48/15 lines increases and peaks around $2,000$~cm$^{-3}$. At 
densities above $10^5$~cm$^{-3}$, the fractional populations of all components
converge to constant values. The trends are similar to Multiplet V3 discussed 
above.

The intensity ratio $I(\lambda5005.15)$/$I(\lambda5001.48 + \lambda5001.14)$
serves as another potential density diagnostic. In reality, however, given the 
closeness in wavelength of the $\lambda$5005.15 line to the [O~{\sc iii}] 
$\lambda$5007 nebular line, which is often several orders of magnitude (3 - 4) 
brighter, accurate measurement of $\lambda$5005.15 line is essentially 
impossible.

\subsubsection{\label{V39}
2s$^2$2p4f\,G[7/2,9/2]$^{\rm e}$ -- 2s$^2$2p3d\,$^3$F$^{\rm o}$ (V39)}

\begin{figure*}[ht]
\centering
   \includegraphics[width=12cm,angle=-90]{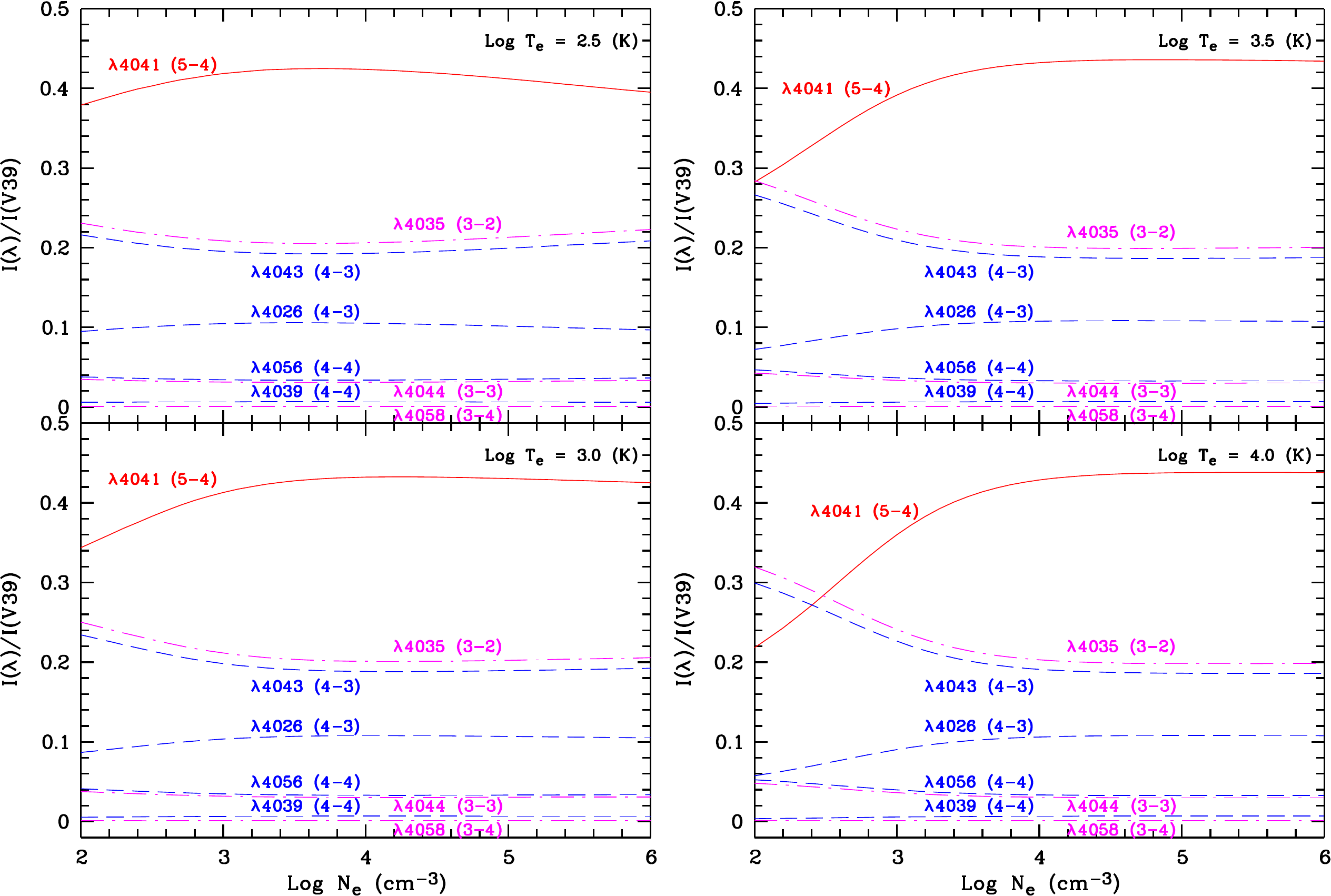}
   \caption{Same as Fig.~\ref{multipletV3} but for Multiplet V39: 
            2s$^2$2p4f\,G[7/2,9/2]$^{\rm e}$ -- 2s$^2$2p3d\,$^3$F$^{\rm o}$.}
   \label{multipletV39}
\end{figure*}

The fractional intensities of fine-structure components of Multiplet V39, 
2s$^2$2p4f\,G[7/2,9/2]$^{\rm e}$ -- 2s$^2$2p3d\,$^3$F$^{\rm o}$, are 
presented in Figure~\ref{multipletV39}. The strongest component 
$\lambda$4041.31 forms exclusively from recombination of target 
$^2$P$^{\rm o}$\,$_{3/2}$ plus cascades from higher states, while the second 
and third strongest components, $\lambda$4035.08 and $\lambda$4043.53, 
which have comparable intensities, can form, in addition, from recombination 
of target $^2$P$^{\rm o}$\,$_{1/2}$.

The behaviour of the intensity of the $\lambda$4041.31 line relative to 
those of the $\lambda$4035.08 and $\lambda$4043.53 lines as a function of 
electron density is quite similar to those of their counterparts of 
Multiplets V3 and V19 discussed above.

The line ratios $I(\lambda4041.31)$/$I(\lambda4035.08)$ and 
$I(\lambda4041.31)$/$I(\lambda4043.53)$ can in principle serve as 
additional density diagnostics. There are however complications in 
their applications:

(1) All fine-structure components of Multiplet V39 
2s$^2$2p4f\,G[7/2,9/2]$^{\rm e}$ -- 2s$^2$2p3d\,$^3$F$^{\rm o}$ are extremely 
faint. The strongest component $\lambda$4041.31 is 2 -- 3 times fainter than 
$\lambda$5679.56, the strongest component of V3, while the latter is 
typically one thousand times fainter than H$\beta$ in a real nebula.

(2) The $\lambda$4041.31 line is blended with the \ion{O}{ii} recombination 
line $\lambda$4041.29 of Multiplet V50c 2p$^2$4f F[2]$^{\rm o}$\,$_{5/2}$ 
-- 2p$^2$3d $^4$F$^{\rm e}$\,$_{5/2}$, while the $\lambda$4035.08 line is 
blended with the \ion{O}{ii} lines $\lambda$4035.07 of Multiplet V50b 
2p$^2$4f F[3]$^{\rm o}$\,$_{5/2}$ -- 2p$^2$3d $^4$F$^{\rm e}$\,$_{5/2}$ 
and $\lambda$4035.49 of Multiplet V50b 2p$^2$4f F[3]$^{\rm o}$\,$_{7/2}$ -- 
2p$^2$3d $^4$F$^{\rm e}$\,$_{5/2}$.

\subsection{\label{s:result3}
Plasma diagnostics}

\begin{figure*}[ht]
\centering
  \includegraphics[width=10cm,angle=-90]{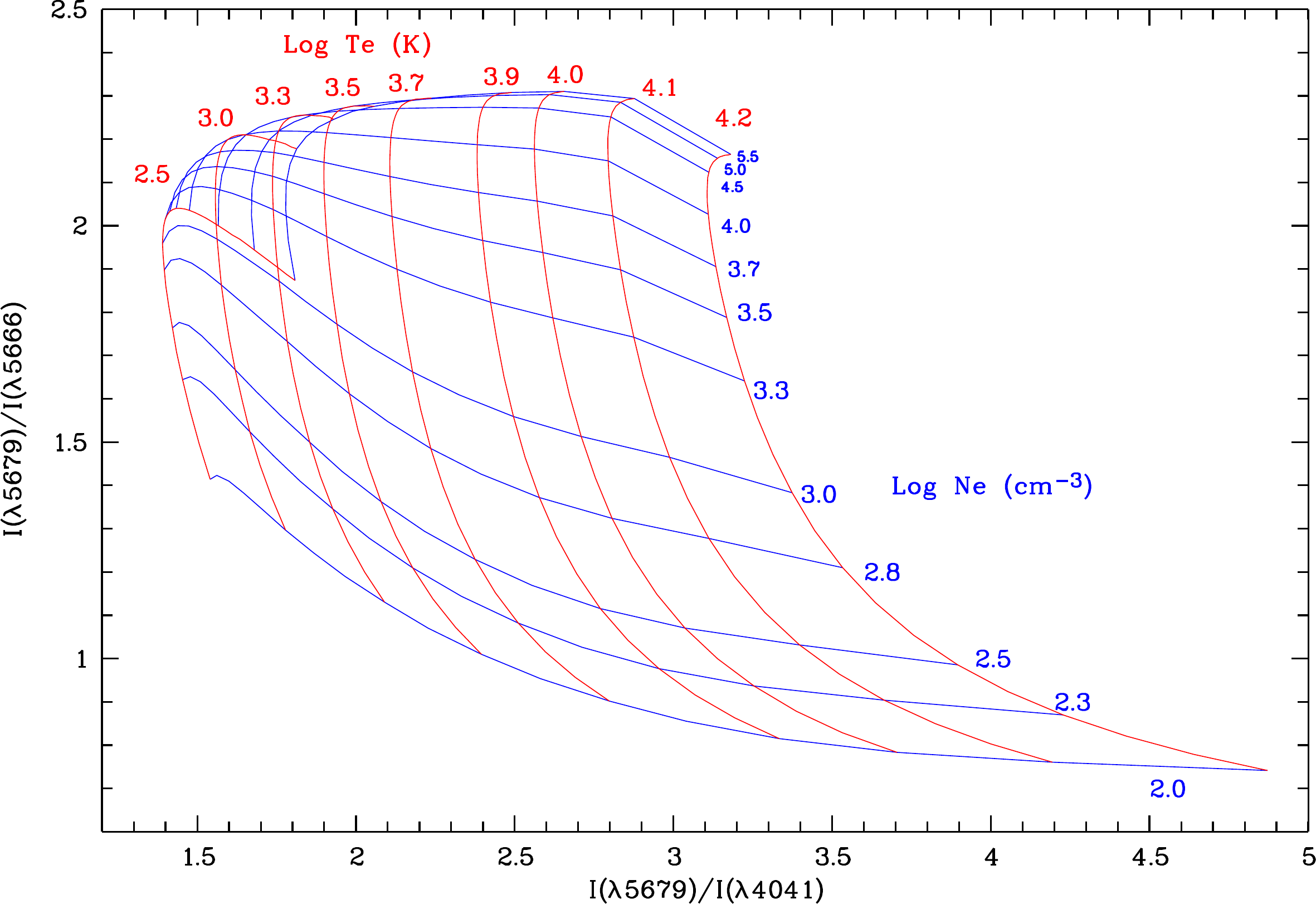}
  \caption{Loci of the \ion{N}{ii} recombination line ratios 
           $I(\lambda5679.56)$/$I(\lambda5666.63)$ and 
           $I(\lambda5679.56)$/$I(\lambda4041.31)$ for different 
           $T_\mathrm{e}$'s and $N_\mathrm{e}$'s.}
  \label{loci}
\end{figure*}

Unlike the UV and optical CELs, whose emissivities have an exponential 
dependence on $T_\mathrm{e}$ (Osterbrock \& Ferland \citealt{OF06}), 
emissivities of heavy element ORLs have only a relatively weak, power-law 
dependence on $T_\mathrm{e}$. The dependence varies for lines originating 
from levels of different orbital angular momentum quantum number $l$. Thus 
the relative intensities of ORLs can also be used to derive electron 
temperature, provided very accurate measurements can be secured (Liu 
\citealt{Liu03}; Liu et al. \citealt{Liu04}; Tsamis et al. 
\cite{tsamis2004}). In the case of \ion{N}{ii}, the intensity ratio of 
$\lambda$5679.56 and $\lambda$4041.31 lines, the strongest components 
of Multiplets V3 3p\,$^3$D$^{\rm e}$ -- 3s\,$^3$P$^{\rm o}$ and V39 
4f\,G[9/2]$^{\rm e}$ -- 3d\,$^3$F$^{\rm o}$, respectively, has a relatively 
strong temperature dependence, and thus can serve as a temperature 
diagnostic. As shown in Section~\ref{s:result2} above, the \ion{N}{ii} 
line ratio $I(\lambda5679.56)$/$I(\lambda5666.63)$ is a good density 
diagnostic. Combining the two line ratios thus allows one to determine
$T_\mathrm{e}$ and $N_\mathrm{e}$ simultaneously. Figure~\ref{loci} shows 
the loci of \ion{N}{ii} recombination line ratios 
$I(\lambda5679.56)$/$I(\lambda5666.63)$ and 
$I(\lambda5679.56)$/$I(\lambda4041.31)$ for different electron temperatures
and densities. With high quality measurements of the two line ratios, one can
readout $T_\mathrm{e}$ and $N_\mathrm{e}$ directly from the diagram.

Ions such as \ion{N}{ii} and \ion{O}{ii} have a rich optical recombination 
line spectrum. Rather than relying on specific line ratios, it is probably 
beneficial and more robust to determine $T_\mathrm{e}$ and $N_\mathrm{e}$ 
by fitting all lines with a good measurement and free from blending 
simultaneously. Details about this approach and its application to 
photoionized gaseous nebulae will be the subject of a subsequent paper.

\subsection{\label{s:result4}
Population of excited states of \ion{N}{$^{2+}$}}

In the current calculations of the effective recombination coefficients of 
\ion{N}{ii}, we have assumed that only the ground fine-structure levels of 
the recombining ion, \ion{N}{$^{2+}$} 2s$^2$2p~$^2$P$^{\rm o}$\,$_{1/2, 3/2}$,
are populated. This is a good approximation under typical nebular conditions. 
The first excited spectral term of N$^{2+}$, 2s2p$^2$~$^4$P, lies 
57\,161.7~cm$^{-1}$ above the ground term (Eriksson \citealt{Eriksson83}), 
and the population of this term is $1.75\times10^{-6}$ relative to that of 
the ground term $^2$P$^{\rm o}$ even at the highest temperature and density 
considered in the current work, $T_\mathrm{e} = 20,000$~K and $N_\mathrm{e} = 
10^6$~cm$^{-3}$. Recombination from the 2s2p$^2$~$^4$P term is thus 
completely negligible.

\subsection{\label{s:result5}
Total recombination coefficients}

Calculations presented in the current work are carried out in 
intermediate coupling in Case B, representing a significant improvement 
compared to Kisielius \& Storey \cite{KS02}, in which the calculations are 
entirely in $LS$-coupling. In our calculations, we have also considered the 
fact that the ground term of \ion{N}{$^{2+}$} comprises two fine-structure 
levels, $^2$P$^{\mathrm{o}}$\,$_{1/2}$ and $^2$P$^{\mathrm{o}}$\,$_{3/2}$, 
and the populations of those two levels deviate from the Boltzmann 
distribution under typical nebular densities. We thus treat the 
recombination of the high-$l$ states using the close-coupling 
photoionization data incorporating the population distribution among 
the $^2$P$^{\mathrm{o}}_{J}$ levels. The critical electron density, at which 
the rates of collisional de-excitation and radiative decay from 
$^2$P$^{\rm o}$\,$_{3/2}$ to $^2$P$^{\rm o}$\,$_{1/2}$ are equal, is 
approximately $2,000$~cm$^{-3}$. At this density, our calculation shows that 
the populations of the $^2$P$^{\rm o}$\,$_{1/2}$ and $^2$P$^{\rm o}$\,$_{3/2}$
levels differ from their Boltzmann values by approximately $34\%$ at 
$T_\mathrm{e}$ = $10,000$~K, The difference given by Kisielius \& Storey 
\cite{KS02} at this density is $30\%$.

In Table~\ref{comparedir}, we compare our direct recombination coefficients 
with those calculated by Nahar \cite{Nahar95} and by Kisielius \& Storey 
\cite{KS02}. The calculations of Nahar \cite{Nahar95} and Kisielius \& Storey 
\cite{KS02} are both in $LS$-coupling, and the \ion{N}{ii} states are not 
$J$-resolved. Their direct recombination coefficients are all to spectral 
term $^{2S+1}L^{\pi}$. Our present calculations are in intermediate coupling,
and recombinations are all to $J$-resolved levels. In order to compare to 
their results, we sum the direct recombination coefficients to all the 
fine-structure levels belonging to individual spectral terms.

At $1,000$~K, the differences between the results of Kisielius \& Storey 
\cite{KS02} and ours are less than $10\%$ for most cases, except for the 
state 2s2p$^3$~$^3$S$^{\rm o}$. For this state, our direct recombination 
coefficient is $50\%$ larger than that of Kisielius \& Storey \cite{KS02}.

At this temperature ($1,000$~K is about 0.1 eV), we believe we have found 
out the exact energy positions for all the resonances below 0.1 Ryd above 
the ionization threshold of \ion{N}{iii} 2p~$^2$P$^{\rm o}$\,$_{1/2}$. 
This region contains most of the important resonances that dominate the 
total recombination rate. In our photoionization calculations, all 
resonances from those of widths as narrow as 10$^{-9}$~Ryd to those of 
widths as wide as 10$^{-4}$~Ryd, are properly resolved using a highly 
adaptive energy mesh. There are typically about 22 points sampling each 
resonance. In the calculation by Kisielius \& Storey \cite{KS02}, the number 
is about ten, while in Nahar \cite{Nahar95} a fixed interval of 0.0004~Ryd 
is used in this energy range.

The three low-lying resonances, $^3$P, $^3$D and $^3$F belonging to the 
2s2p$^2$($^4$P)\,3d configuration, are situated between 0.075 and 0.085 Ryd 
above the ionization threshold of $^2$P$^{\rm o}$\,$_{1/2}$ (Kisielius \& 
Storey \citealt{KS02}). For the state 2s2p$^3$\,$^3$S$^{\rm o}$, one of the 
main sources of recombination is from the term $^3$P belonging to the 
2s2p$^2$($^4$P)\,3d configuration.  There are three fine-structure resonance 
levels of the $^3$P term with quantum numbers $J$ = 0, 1 and 2. The full 
widths of these three resonances are $1.02\times10^{-4}$~Ryd for 
$^3$P$^{\rm e}_{0}$, $1.35\times10^{-5}$~Ryd for $^3$P$^{\rm e}_{1}$ and
$1.42\times10^{-5}$~Ryd for $^3$P$^{\rm e}_{2}$. The steps of energy mesh 
adopted for the three resonances are: $4.64\times10^{-6}$~Ryd for 
$^3$P$^{\rm e}_{0}$, $6.14\times10^{-7}$~Ryd for $^3$P$^{\rm e}_{1}$ and 
$6.46\times10^{-7}$~Ryd for $^3$P$^{\rm e}_{2}$.

Our calculations are carried out entirely in intermediate coupling. This 
leads to a high recombination rate to the 2s2p$^3$\,$^3$S$^{\rm o}$ state, 
produced by radiative intercombination transitions (transitions between 
levels of different total spins) from levels above the ionization threshold 
to the 2s2p$^3$\,$^3$S$^{\rm o}$ level. The widths of such intercombination 
transitions are usually much narrower than those of allowed transitions. 
For example, the resonance level $^5$P$^{\rm e}$\,$_{1}$ belonging to the 
configuration 2s2p$^2$($^4$P)\,3d lies about 0.051~Ryd above the ionization 
threshold, and it can decay to the level 2s2p$^3$\,$^3$S$^{\rm o}$\,$_{1}$ 
via an intercombination transition. The width of this resonance is 
$2.49\times10^{-9}$~Ryd, and the energy interval of the photoionization mesh 
is set to $1.13\times10^{-10}$~Ryd. Intercombination transitions were not 
considered in Kisielius \& Storey \cite{KS02}, given the calculations 
were in $LS$-coupling.

At $1,000$~K, the differences between the calculations of Nahar 
\cite{Nahar95} and ours are smaller than $10\%$, except for states belonging 
to the 2s2p$^3$ configuration.

At $10,000$~K, the differences between the calculations of Kisielius \& 
Storey \cite{KS02} and ours are all better than $10\%$. The agreement for 
the state 2s2p$^3$\,$^3$S$^{\rm o}$ is particularly good.

At this temperature, the differences between the results of Nahar 
\cite{Nahar95} and ours are within $15\%$, except for states 
$^3$S$^{\rm o}$, $^3$P$^{\rm o}$ and $^3$D$^{\rm o}$ belonging to the 
configuration 2s2p$^3$, where the differences are larger than $30\%$. The 
large discrepancies are likely to be caused by the coarse energy mesh adopted 
by Nahar \cite{Nahar95} for the photoionization calculations, leading to the 
recombination rates to states belonging to the 2s2p$^3$ configuration being 
underestimated.

In Table~\ref{comparedir}, we compare our total direct recombination 
coefficients, which are the sum of all the direct recombination coefficients 
to individual atomic levels with $n \leq 35$, with those of Nahar 
\cite{Nahar95} and Kisielius \& Storey \cite{KS02}. At $1,000$~K, our total 
recombination coefficient is 13 per cent lower than that of Kisielius \& 
Storey \cite{KS02}. That is probably because the sum only reaches up to 
$n = 35$. At $10,000$~K, our total recombination coefficient is higher than 
the other two.

\addtocounter{table}{1}

\section{\label{s:conclusion}
Conclusion}

Effective recombination coefficients for the \ion{N}{$^+$} recombination 
line spectrum have been calculated in Case B for a wide range of electron 
density and temperature. The results are fitted with analytical formulae 
as a function of electron temperature for different electron densities, 
to an accuracy of better than 0.5\%.

The high quality basic atomic data adopted in the current work, including 
photoionization cross-sections, bound-bound transition probabilities, and 
bound state energy values, were obtained from R-matrix calculations 
for all bound states with $n \leq 11$ in the intermediate coupling scheme. 
All major resonances near the ionization thresholds were properly 
resolved. In calculating the \ion{N}{ii} level populations, we took 
into account the fact that the populations of the ground fine-structure 
levels of the recombining ion \ion{N}{$^{2+}$} deviate from the Boltzmann 
distribution. Fine-structure dielectronic recombination, which occurs through
high Rydberg states lying between the doublet $^2$P$^{\rm o}_{1/2,\,3/2}$ 
thresholds and is very effective at low temperatures ($\leq 250$~K), 
was also included in the current investigation. The calculations 
extend to $l \leq 4$.

The effective recombination coefficients for the \ion{N}{ii} 
recombination spectrum presented in the current work represent 
recombination processes under typical nebular conditions. The
sensitivity of individual lines within a multiplet to the density and
temperature of the emitting medium opens up the possibility of electron
temperature and density diagnostics and abundance determinations which
were not possible with earlier theory.

%

\tiny
\longtab{3}{
\begin{longtable}{lrlccccccc}
\caption{\label{aeff_Ne100} Case~B effective recombination coefficients [$10^{-15}$ cm$^{3}$~s$^{-1}$] for electron density $N_\mathrm{e}$ = 10$^2$~cm$^{-3}$.}\\
\cline{1-10}\\
& & & \multicolumn{7}{c}{$T_{\mathrm e}$\,[K]}\\
\cline{4-10}\\
Transition& $\lambda$[\AA] & & 125 & 500 & 1000 & 5000 & 10,000 & 15,000 & 20,000 \\
\cline{1-10}\\
\endfirsthead

\caption{Continued.}\\
\cline{1-10}\\
& & & \multicolumn{7}{c}{$T_{\mathrm e}$\,[K]}\\
\cline{4-10}\\
Transition& $\lambda$[\AA] & & 125 & 500 & 1000 & 5000 & 10,000 & 15,000 & 20,000 \\
\cline{1-10}\\
\endhead
\hline
\endfoot

 6h -- 5g & & & & & & & & \\
2p6h\,$I[13/2]^{\rm e}_{6}$ -- 2p5g\,$H[11/2]^{\rm o}_{5}$    & 18622.57 &  &  1.54e$+$3   &  3.44e$+$2   &  1.43e$+$2   &  1.47e$+$1   &  5.24e$+$0   &   2.88e$+$0  &  1.84e$+$0\\
2p6h\,$H[11/2]^{\rm e}_{6}$ -- 2p5g\,$G[9/2]^{\rm o}_{5}$     & 18481.01 &  &  1.71e$+$3   &  3.75e$+$2   &  1.55e$+$2   &  1.66e$+$1   &  6.01e$+$0   &   3.33e$+$0  &  2.14e$+$0\\
2p6h\,$H[11/2]^{\rm e}_{5}$ -- 2p5g\,$G[9/2]^{\rm o}_{4}$     & 17964.67 &  &  1.42e$+$3   &  3.12e$+$2   &  1.29e$+$2   &  1.38e$+$1   &  5.00e$+$0   &   2.76e$+$0  &  1.78e$+$0\\
2p6h\,$H[9/2]^{\rm e}_{5}$ -- 2p5g\,$G[7/2]^{\rm o}_{4}$      & 18480.67 &  &  1.37e$+$3   &  3.02e$+$2   &  1.25e$+$2   &  1.32e$+$1   &  4.80e$+$0   &   2.65e$+$0  &  1.70e$+$0\\
2p6h\,$G[9/2]^{\rm e}_{4}$ -- 2p5g\,$F[7/2]^{\rm o}_{3}$      & 18549.42 &  &  1.09e$+$3   &  2.40e$+$2   &  9.94e$+$1   &  1.05e$+$1   &  3.82e$+$0   &   2.11e$+$0  &  1.36e$+$0\\
2p6h\,$G[7/2]^{\rm e}_{4}$ -- 2p5g\,$F[5/2]^{\rm o}_{3}$      & 18679.36 &  &  9.71e$+$2   &  2.18e$+$2   &  9.02e$+$1   &  9.19e$+$0   &  3.24e$+$0   &   1.78e$+$0  &  1.14e$+$0\\
2p6h\,$I[13/2]^{\rm e}_{7}$ -- 2p5g\,$H[11/2]^{\rm o}_{6}$    & 18622.64 &  &  1.80e$+$3   &  4.02e$+$2   &  1.67e$+$2   &  1.72e$+$1   &  6.12e$+$0   &   3.36e$+$0  &  2.15e$+$0\\
\end{longtable}
}

\tiny
\longtab{4}{
\begin{longtable}{lrlccccccc}
\caption{\label{aeff_Ne1000} Same as Table~\ref{aeff_Ne100} but for $N_\mathrm{e}$ = 10$^3$~cm$^{-3}$.}\\
\cline{1-10}\\
& & & \multicolumn{7}{c}{$T_{\mathrm e}$\,[K]}\\
\cline{4-10}\\
Transition& $\lambda$[\AA] & & 125 & 500 & 1000 & 5000 & 10,000 & 15,000 & 20,000 \\
\cline{1-10}\\
\endfirsthead

\caption{Continued.}\\
\cline{1-10}\\
& & & \multicolumn{7}{c}{$T_{\mathrm e}$\,[K]}\\
\cline{4-10}\\
Transition& $\lambda$[\AA] & & 125 & 500 & 1000 & 5000 & 10,000 & 15,000 & 20,000 \\
\cline{1-10}\\
\endhead
\hline
\endfoot

 6h -- 5g & & & & & & & & \\
2p6h\,$I[13/2]^{\rm e}_{6}$ -- 2p5g\,$H[11/2]^{\rm o}_{5}$    & 18622.57 &  &  1.60e$+$3   &  3.30e$+$2   &  1.47e$+$2   &  1.95e$+$1   &  7.54e$+$0   &  4.27e$+$0   &  2.78e$+$0\\
2p6h\,$H[11/2]^{\rm e}_{6}$ -- 2p5g\,$G[9/2]^{\rm o}_{5}$     & 18481.01 &  &  1.59e$+$3   &  3.23e$+$2   &  1.44e$+$2   &  1.91e$+$1   &  7.41e$+$0   &  4.21e$+$0   &  2.75e$+$0\\
2p6h\,$H[11/2]^{\rm e}_{5}$ -- 2p5g\,$G[9/2]^{\rm o}_{4}$     & 17964.67 &  &  1.32e$+$3   &  2.69e$+$2   &  1.19e$+$2   &  1.59e$+$1   &  6.16e$+$0   &  3.50e$+$0   &  2.28e$+$0\\
2p6h\,$H[9/2]^{\rm e}_{5}$ -- 2p5g\,$G[7/2]^{\rm o}_{4}$      & 18480.67 &  &  1.29e$+$3   &  2.63e$+$2   &  1.17e$+$2   &  1.56e$+$1   &  6.04e$+$0   &  3.43e$+$0   &  2.23e$+$0\\
2p6h\,$G[9/2]^{\rm e}_{4}$ -- 2p5g\,$F[7/2]^{\rm o}_{3}$      & 18549.42 &  &  1.03e$+$3   &  2.10e$+$2   &  9.32e$+$1   &  1.24e$+$1   &  4.80e$+$0   &  2.73e$+$0   &  1.78e$+$0\\
2p6h\,$G[7/2]^{\rm e}_{4}$ -- 2p5g\,$F[5/2]^{\rm o}_{3}$      & 18679.36 &  &  1.03e$+$3   &  2.12e$+$2   &  9.46e$+$1   &  1.26e$+$1   &  4.85e$+$0   &  2.74e$+$0   &  1.78e$+$0\\
2p6h\,$I[13/2]^{\rm e}_{7}$ -- 2p5g\,$H[11/2]^{\rm o}_{6}$    & 18622.64 &  &  1.87e$+$3   &  3.85e$+$2   &  1.72e$+$2   &  2.28e$+$1   &  8.81e$+$0   &  4.99e$+$0   &  3.25e$+$0\\
\end{longtable}
}

\tiny
\longtab{5}{
\begin{longtable}{lrlccccccc}
\caption{\label{aeff_Ne10000} Same as Table~\ref{aeff_Ne100} but for $N_\mathrm{e}$ = 10$^4$~cm$^{-3}$.}\\
\cline{1-10}\\
& & & \multicolumn{7}{c}{$T_{\mathrm e}$\,[K]}\\
\cline{4-10}\\
Transition& $\lambda$[\AA] & & 125 & 500 & 1000 & 5000 & 10,000 & 15,000 & 20,000 \\
\cline{1-10}\\
\endfirsthead

\caption{Continued.}\\
\cline{1-10}\\
& & & \multicolumn{7}{c}{$T_{\mathrm e}$\,[K]}\\
\cline{4-10}\\
Transition& $\lambda$[\AA] & & 125 & 500 & 1000 & 5000 & 10,000 & 15,000 & 20,000 \\
\cline{1-10}\\
\endhead
\hline
\endfoot

 6h -- 5g & & & & & & & & \\
2p6h\,$I[13/2]^{\rm e}_{6}$ -- 2p5g\,$H[11/2]^{\rm o}_{5}$    & 18622.57 &  &  1.34e$+$3   &  2.81e$+$2   &  1.32e$+$2   &  2.06e$+$1   &  8.53e$+$0   &  5.01e$+$0   &  3.35e$+$0\\
2p6h\,$H[11/2]^{\rm e}_{6}$ -- 2p5g\,$G[9/2]^{\rm o}_{5}$     & 18481.01 &  &  1.29e$+$3   &  2.70e$+$2   &  1.26e$+$2   &  1.97e$+$1   &  8.17e$+$0   &  4.81e$+$0   &  3.21e$+$0\\
2p6h\,$H[11/2]^{\rm e}_{5}$ -- 2p5g\,$G[9/2]^{\rm o}_{4}$     & 17964.67 &  &  1.07e$+$3   &  2.24e$+$2   &  1.05e$+$2   &  1.64e$+$1   &  6.79e$+$0   &  3.99e$+$0   &  2.67e$+$0\\
2p6h\,$H[9/2]^{\rm e}_{5}$ -- 2p5g\,$G[7/2]^{\rm o}_{4}$      & 18480.67 &  &  1.05e$+$3   &  2.20e$+$2   &  1.03e$+$2   &  1.61e$+$1   &  6.68e$+$0   &  3.92e$+$0   &  2.63e$+$0\\
2p6h\,$G[9/2]^{\rm e}_{4}$ -- 2p5g\,$F[7/2]^{\rm o}_{3}$      & 18549.42 &  &  8.37e$+$2   &  1.75e$+$2   &  8.21e$+$1   &  1.28e$+$1   &  5.31e$+$0   &  3.12e$+$0   &  2.09e$+$0\\
2p6h\,$G[7/2]^{\rm e}_{4}$ -- 2p5g\,$F[5/2]^{\rm o}_{3}$      & 18679.36 &  &  8.61e$+$2   &  1.81e$+$2   &  8.46e$+$1   &  1.32e$+$1   &  5.48e$+$0   &  3.22e$+$0   &  2.16e$+$0\\
2p6h\,$I[13/2]^{\rm e}_{7}$ -- 2p5g\,$H[11/2]^{\rm o}_{6}$    & 18622.64 &  &  1.56e$+$3   &  3.28e$+$2   &  1.54e$+$2   &  2.40e$+$1   &  9.96e$+$0   &  5.86e$+$0   &  3.92e$+$0\\
\end{longtable}
}

\tiny
\longtab{6}{
\begin{longtable}{lrlccccccc}
\caption{\label{aeff_Ne100000} Same as Table~\ref{aeff_Ne100} but for $N_\mathrm{e}$ = 10$^5$~cm$^{-3}$.}\\
\cline{1-10}\\
& & & \multicolumn{7}{c}{$T_{\mathrm e}$\,[K]}\\
\cline{4-10}
Transition& $\lambda$[\AA] & & 125 & 500 & 1000 & 5000 & 10,000 & 15,000 & 20,000 \\
\cline{1-10}\\
\endfirsthead

\caption{Continued.}\\
\cline{1-10}\\
& & & \multicolumn{7}{c}{$T_{\mathrm e}$\,[K]}\\
\cline{4-10}\\
Transition& $\lambda$[\AA] & & 125 & 500 & 1000 & 5000 & 10,000 & 15,000 & 20,000 \\
\cline{1-10}\\
\endhead
\hline
\endfoot

 6h -- 5g & & & & & & & & \\
2p6h\,$I[13/2]^{\rm e}_{6}$ -- 2p5g\,$H[11/2]^{\rm o}_{5}$    & 18622.57 &  &  1.03e$+$0   &  2.26e$+$2   &  1.10e$+$2   &  1.89e$+$1   &  8.14e$+$0   &  4.87e$+$0   &  3.30e$+$0\\
2p6h\,$H[11/2]^{\rm e}_{6}$ -- 2p5g\,$G[9/2]^{\rm o}_{5}$     & 18481.01 &  &  9.82e$+$2   &  2.17e$+$2   &  1.05e$+$2   &  1.81e$+$1   &  7.79e$+$0   &  4.67e$+$0   &  3.16e$+$0\\
2p6h\,$H[11/2]^{\rm e}_{5}$ -- 2p5g\,$G[9/2]^{\rm o}_{4}$     & 17964.67 &  &  8.16e$+$2   &  1.80e$+$2   &  8.76e$+$1   &  1.50e$+$1   &  6.47e$+$0   &  3.88e$+$0   &  2.63e$+$0\\
2p6h\,$H[9/2]^{\rm e}_{5}$ -- 2p5g\,$G[7/2]^{\rm o}_{4}$      & 18480.67 &  &  8.02e$+$2   &  1.77e$+$2   &  8.61e$+$1   &  1.48e$+$1   &  6.36e$+$0   &  3.81e$+$0   &  2.58e$+$0\\
2p6h\,$G[9/2]^{\rm e}_{4}$ -- 2p5g\,$F[7/2]^{\rm o}_{3}$      & 18549.42 &  &  6.38e$+$2   &  1.41e$+$2   &  6.85e$+$1   &  1.18e$+$1   &  5.06e$+$0   &  3.03e$+$0   &  2.06e$+$0\\
2p6h\,$G[7/2]^{\rm e}_{4}$ -- 2p5g\,$F[5/2]^{\rm o}_{3}$      & 18679.36 &  &  6.59e$+$2   &  1.45e$+$2   &  7.07e$+$1   &  1.22e$+$1   &  5.23e$+$0   &  3.13e$+$0   &  2.12e$+$0\\
2p6h\,$I[13/2]^{\rm e}_{7}$ -- 2p5g\,$H[11/2]^{\rm o}_{6}$    & 18622.64 &  &  1.20e$+$3   &  2.64e$+$2   &  1.29e$+$2   &  2.21e$+$1   &  9.51e$+$0   &  5.70e$+$0   &  3.86e$+$0\\
\end{longtable}
}

\longtab{7}{
\begin{longtable}{llrrrrrrr}
\caption[]{\label{fit1_lowTe} Fit parameters and maximum errors $\delta$[\%] for $T_\mathrm{e} < 10,000$~K and $N_\mathrm{e} = 10^2$~cm$^{-3}$. The results are for Case B recombination.}\\
\cline{1-9}\\
Transition & $\lambda$[\AA] & $a_0$ & $a_1$ & $a_2$ & $a_3$ & $a_4$ & $a_5$ & $\delta[\%]$ \\
\cline{1-9}\\
\endfirsthead

\caption{Continued.}\\
\cline{1-9}\\
Transition & $\lambda$[\AA] & $a_0$ & $a_1$ & $a_2$ & $a_3$ & $a_4$ & $a_5$ & $\delta[\%]$ \\
\cline{1-9}\\
\endhead
\hline
\endfoot

2p3d($^3$F$^{\rm o}_{2}$) -- 2p3p($^3$D$^{\rm e}_{1}$)  &   5001.14  &   22.7408  &  -26.4693  &   15.2874  &   -4.6280  &    0.7091  &   -0.0439  &  0.356\\
2p3d($^3$F$^{\rm o}_{3}$) -- 2p3p($^3$D$^{\rm e}_{2}$)  &   5001.48  &   27.4981  &  -33.7298  &   19.9870  &   -6.1467  &    0.9514  &   -0.0591  &  0.543\\
2p3d($^3$F$^{\rm o}_{4}$) -- 2p3p($^3$D$^{\rm e}_{3}$)  &   5005.15  &   -1.1211  &    9.8186  &   -6.6487  &    2.0583  &   -0.3243  &    0.0206  &  0.361\\
2p3d($^3$F$^{\rm o}_{2}$) -- 2p3p($^3$D$^{\rm e}_{2}$)  &   5016.39  &   22.2658  &  -26.9751  &   15.6171  &   -4.7344  &    0.7262  &   -0.0449  &  0.487\\
2p3d($^3$F$^{\rm o}_{3}$) -- 2p3p($^3$D$^{\rm e}_{3}$)  &   5025.66  &   26.5231  &  -33.7292  &   19.9662  &   -6.1334  &    0.9483  &   -0.0588  &  0.454\\
2p3d($^3$F$^{\rm o}_{2}$) -- 2p3p($^3$D$^{\rm e}_{3}$)  &   5040.72  &   20.2540  &  -26.3467  &   15.2021  &   -4.5986  &    0.7042  &   -0.0435  &  0.317\\
2p3d($^3$D$^{\rm o}_{2}$) -- 2p3p($^3$D$^{\rm e}_{1}$)  &   4774.24  &   17.6233  &  -23.3485  &   13.2635  &   -3.9552  &    0.5963  &   -0.0364  &  0.316\\
2p3d($^3$D$^{\rm o}_{1}$) -- 2p3p($^3$D$^{\rm e}_{1}$)  &   4779.72  &   28.7969  &  -36.1782  &   20.8559  &   -6.2078  &    0.9318  &   -0.0564  &  0.399\\
\hline
\end{longtable}}

\longtab{8}{
\begin{longtable}{llrrrrrrrr}
\caption[]{\label{fit1_highTe} The same as Table~\ref{fit1_lowTe} but for $10,000 \leq T_\mathrm{e} \leq 20,000$~K and $N_\mathrm{e} = 10^2$~cm$^{-3}$.}\\
\cline{1-10}\\
Transition & $\lambda$[\AA] & $b_0$ & $b_1$ & $b_2$ & $b_3$ & $b_4$ & $b_5$ & $b_6$ & $\delta[\%]$ \\
\cline{1-10}\\
\endfirsthead

\caption{Continued.}\\
\cline{1-10}\\
Transition & $\lambda$[\AA] & $b_0$ & $b_1$ & $b_2$ & $b_3$ & $b_4$ & $b_5$ & $b_6$ & $\delta[\%]$ \\
\cline{1-10}\\
\endhead
\hline
\endfoot

2p3d($^3$F$^{\rm o}_{2}$) -- 2p3p($^3$D$^{\rm e}_{1}$)  &   5001.14  &    4.0807  &   -4.8537  &   10.8131  &   -5.7467  &    1.6643  &   -0.0937  &   -1.1401  &  0.105\\ 
2p3d($^3$F$^{\rm o}_{3}$) -- 2p3p($^3$D$^{\rm e}_{2}$)  &   5001.48  &    8.5638  &   -9.2646  &   12.9159  &   -6.1298  &    1.6024  &    0.5439  &   -1.3349  &  0.112\\
2p3d($^3$F$^{\rm o}_{4}$) -- 2p3p($^3$D$^{\rm e}_{3}$)  &   5005.15  &    1.3006  &   -1.3424  &    5.5162  &   -2.7575  &    0.8097  &   -0.8554  &   -0.8026  &  0.118\\
2p3d($^3$F$^{\rm o}_{2}$) -- 2p3p($^3$D$^{\rm e}_{2}$)  &   5016.39  &   12.9496  &  -20.2178  &   21.2091  &   -9.9809  &    2.2035  &    1.1948  &   -1.7026  &  0.065\\
2p3d($^3$F$^{\rm o}_{3}$) -- 2p3p($^3$D$^{\rm e}_{3}$)  &   5025.66  &    7.8221  &  -11.9792  &   14.4626  &   -7.3337  &    1.7237  &    0.7413  &   -1.5171  &  0.019\\
2p3d($^3$F$^{\rm o}_{2}$) -- 2p3p($^3$D$^{\rm e}_{3}$)  &   5040.72  &   -7.1739  &    9.6530  &   -7.4632  &    2.3748  &   -0.2830  &    2.7094  &   -1.7328  &  0.013\\
2p3d($^3$D$^{\rm o}_{2}$) -- 2p3p($^3$D$^{\rm e}_{1}$)  &   4774.24  &   -7.3956  &    7.2197  &   -6.1707  &    1.7555  &   -0.2650  &    1.4554  &   -1.3196  &  0.126\\
2p3d($^3$D$^{\rm o}_{1}$) -- 2p3p($^3$D$^{\rm e}_{1}$)  &   4779.72  &    0.7178  &    0.4617  &    1.9964  &   -0.9810  &    0.3059  &   -0.7224  &   -0.7097  &  0.138\\
\hline
\end{longtable}}

\longtab{9}{
\begin{longtable}{llrrrrrrr}
\caption[]{\label{fit2_lowTe} The same as Table~\ref{fit1_lowTe} but for $T_\mathrm{e} < 10,000$~K and $N_\mathrm{e} = 10^3$~cm$^{-3}$.}\\
\cline{1-9}\\
Transition & $\lambda$[\AA] & $a_0$ & $a_1$ & $a_2$ & $a_3$ & $a_4$ & $a_5$ & $\delta[\%]$ \\
\cline{1-9}\\
\endfirsthead

\caption{Continued.}\\
\cline{1-9}\\
Transition & $\lambda$[\AA] & $a_0$ & $a_1$ & $a_2$ & $a_3$ & $a_4$ & $a_5$ & $\delta[\%]$ \\
\cline{1-9}\\
\endhead
\hline
\endfoot

2p3d($^3$F$^{\rm o}_{2}$) -- 2p3p($^3$D$^{\rm e}_{1}$)  &   5001.14  &   12.5791  &   -9.3536  &    3.8787  &   -0.9076  &    0.1131  &   -0.0062  &  0.147\\
2p3d($^3$F$^{\rm o}_{3}$) -- 2p3p($^3$D$^{\rm e}_{2}$)  &   5001.48  &   16.1471  &  -14.7117  &    7.2411  &   -1.9366  &    0.2665  &   -0.0151  &  0.258\\
2p3d($^3$F$^{\rm o}_{4}$) -- 2p3p($^3$D$^{\rm e}_{3}$)  &   5005.15  &   -5.3569  &   18.9374  &  -13.7814  &    4.6487  &   -0.7664  &    0.0495  &  0.420\\
2p3d($^3$F$^{\rm o}_{2}$) -- 2p3p($^3$D$^{\rm e}_{2}$)  &   5016.39  &   11.9406  &   -9.5913  &    4.0310  &   -0.9549  &    0.1202  &   -0.0066  &  0.177\\
2p3d($^3$F$^{\rm o}_{3}$) -- 2p3p($^3$D$^{\rm e}_{3}$)  &   5025.66  &   14.9808  &  -14.3695  &    6.9773  &   -1.8372  &    0.2482  &   -0.0138  &  0.161\\
2p3d($^3$F$^{\rm o}_{2}$) -- 2p3p($^3$D$^{\rm e}_{3}$)  &   5040.72  &   10.5551  &  -10.0243  &    4.3291  &   -1.0564  &    0.1373  &   -0.0077  &  0.164\\
2p3d($^3$D$^{\rm o}_{2}$) -- 2p3p($^3$D$^{\rm e}_{1}$)  &   4774.24  &    8.0000  &   -6.9925  &    2.3368  &   -0.4038  &    0.0315  &   -0.0010  &  0.126\\
2p3d($^3$D$^{\rm o}_{1}$) -- 2p3p($^3$D$^{\rm e}_{1}$)  &   4779.72  &   17.8006  &  -18.0320  &    8.9852  &   -2.4130  &    0.3362  &   -0.0195  &  0.208\\
\hline
\end{longtable}}

\longtab{10}{
\begin{longtable}{llrrrrrrrr}
\caption[]{\label{fit2_highTe} The same as Table~\ref{fit1_lowTe} but for $10,000 \leq T_\mathrm{e} \leq 20,000$~K and $N_\mathrm{e} = 10^3$~cm$^{-3}$.}\\
\cline{1-10}\\
Transition & $\lambda$[\AA] & $b_0$ & $b_1$ & $b_2$ & $b_3$ & $b_4$ & $b_5$ & $b_6$ & $\delta[\%]$ \\
\cline{1-10}\\
\endfirsthead

\caption{Continued.}\\
\cline{1-10}\\
Transition & $\lambda$[\AA] & $b_0$ & $b_1$ & $b_2$ & $b_3$ & $b_4$ & $b_5$ & $b_6$ & $\delta[\%]$ \\
\cline{1-10}\\
\endhead
\hline
\endfoot

2p3d($^3$F$^{\rm o}_{2}$) -- 2p3p($^3$D$^{\rm e}_{1}$)  &   5001.14  &    5.0900  &   -9.6681  &   18.5895  &  -10.8190  &    3.1004  &   -0.1127  &   -1.2715  &  0.117\\
2p3d($^3$F$^{\rm o}_{3}$) -- 2p3p($^3$D$^{\rm e}_{2}$)  &   5001.48  &    8.1496  &  -11.1418  &   17.5855  &   -9.3308  &    2.5425  &    0.4062  &   -1.3962  &  0.142\\
2p3d($^3$F$^{\rm o}_{4}$) -- 2p3p($^3$D$^{\rm e}_{3}$)  &   5005.15  &    0.4957  &   -1.9574  &    4.0614  &   -0.7921  &    0.3277  &   -2.5045  &   -0.1236  &  0.097\\
2p3d($^3$F$^{\rm o}_{2}$) -- 2p3p($^3$D$^{\rm e}_{2}$)  &   5016.39  &   10.0017  &  -18.5348  &   24.1598  &  -13.5435  &    3.3781  &    0.8687  &   -1.7156  &  0.056\\
2p3d($^3$F$^{\rm o}_{3}$) -- 2p3p($^3$D$^{\rm e}_{3}$)  &   5025.66  &    8.8667  &  -15.9365  &   20.3392  &  -11.1045  &    2.6920  &    0.7884  &   -1.6433  &  0.028\\
2p3d($^3$F$^{\rm o}_{2}$) -- 2p3p($^3$D$^{\rm e}_{3}$)  &   5040.72  &  -14.9946  &   23.3356  &  -19.0802  &    6.3036  &   -0.7802  &    2.8050  &   -2.0791  &  0.033\\
2p3d($^3$D$^{\rm o}_{2}$) -- 2p3p($^3$D$^{\rm e}_{1}$)  &   4774.24  &   -6.8546  &    7.5362  &   -6.2725  &    2.0281  &   -0.2554  &    1.2853  &   -0.9936  &  0.118\\
2p3d($^3$D$^{\rm o}_{1}$) -- 2p3p($^3$D$^{\rm e}_{1}$)  &   4779.72  &   10.2691  &  -15.2839  &   16.3095  &   -7.6241  &    1.6801  &    1.0306  &   -1.6136  &  0.066\\
\hline
\end{longtable}}

\longtab{11}{
\begin{longtable}{llrrrrrrr}
\caption[]{\label{fit3_lowTe} The same as Table~\ref{fit1_lowTe} but for $T_\mathrm{e} < 10,000$~K and $N_\mathrm{e} = 10^4$~cm$^{-3}$.}\\
\cline{1-9}\\
Transition & $\lambda$[\AA] & $a_0$ & $a_1$ & $a_2$ & $a_3$ & $a_4$ & $a_5$ & $\delta[\%]$ \\
\cline{1-9}\\
\endfirsthead

\caption{Continued.}\\
\cline{1-9}\\
Transition & $\lambda$[\AA] & $a_0$ & $a_1$ & $a_2$ & $a_3$ & $a_4$ & $a_5$ & $\delta[\%]$ \\
\cline{1-9}\\
\endhead
\hline
\endfoot

2p3d($^3$F$^{\rm o}_{2}$) -- 2p3p($^3$D$^{\rm e}_{1}$)  &   5001.14  &   11.7090  &   -6.5834  &    1.3517  &    0.1004  &   -0.0749  &    0.0072  &  0.152\\
2p3d($^3$F$^{\rm o}_{3}$) -- 2p3p($^3$D$^{\rm e}_{2}$)  &   5001.48  &   15.2174  &  -12.0798  &    4.9234  &   -1.0327  &    0.1019  &   -0.0037  &  0.134\\
2p3d($^3$F$^{\rm o}_{4}$) -- 2p3p($^3$D$^{\rm e}_{3}$)  &   5005.15  &   -3.6703  &   16.6101  &  -12.4552  &    4.2428  &   -0.6991  &    0.0449  &  0.323\\
2p3d($^3$F$^{\rm o}_{2}$) -- 2p3p($^3$D$^{\rm e}_{2}$)  &   5016.39  &   11.4997  &   -7.5496  &    1.9948  &   -0.1110  &   -0.0406  &    0.0050  &  0.143\\
2p3d($^3$F$^{\rm o}_{3}$) -- 2p3p($^3$D$^{\rm e}_{3}$)  &   5025.66  &   14.4619  &  -12.4357  &    5.1295  &   -1.0900  &    0.1095  &   -0.0041  &  0.153\\
2p3d($^3$F$^{\rm o}_{2}$) -- 2p3p($^3$D$^{\rm e}_{3}$)  &   5040.72  &    9.7249  &   -7.2919  &    1.8114  &   -0.0475  &   -0.0513  &    0.0057  &  0.070\\
2p3d($^3$D$^{\rm o}_{2}$) -- 2p3p($^3$D$^{\rm e}_{1}$)  &   4774.24  &    7.2931  &   -4.5648  &    0.0977  &    0.4872  &   -0.1341  &    0.0108  &  0.093\\
2p3d($^3$D$^{\rm o}_{1}$) -- 2p3p($^3$D$^{\rm e}_{1}$)  &   4779.72  &   15.4504  &  -12.8741  &    4.9683  &   -0.9514  &    0.0796  &   -0.0019  &  0.223\\
\hline                                                    
\end{longtable}}

\longtab{12}{
\begin{longtable}{llrrrrrrrr}
\caption[]{\label{fit3_highTe} The same as Table~\ref{fit1_lowTe} but for $10,000 \leq T_\mathrm{e} \leq 20,000$~K and $N_\mathrm{e} = 10^4$~cm$^{-3}$.}\\
\cline{1-10}\\
Transition & $\lambda$[\AA] & $b_0$ & $b_1$ & $b_2$ & $b_3$ & $b_4$ & $b_5$ & $b_6$ & $\delta[\%]$ \\
\cline{1-10} \\
\endfirsthead

\caption{Continued.}\\
\cline{1-10}\\
Transition & $\lambda$[\AA] & $b_0$ & $b_1$ & $b_2$ & $b_3$ & $b_4$ & $b_5$ & $b_6$ & $\delta[\%]$ \\
\cline{1-10}\\
\endhead
\hline
\endfoot

2p3d($^3$F$^{\rm o}_{2}$) -- 2p3p($^3$D$^{\rm e}_{1}$)  &   5001.14  &    7.5464  &  -20.6491  &   35.3950  &  -21.6467  &    5.9869  &   -0.2129  &   -1.3891  &  0.171\\
2p3d($^3$F$^{\rm o}_{3}$) -- 2p3p($^3$D$^{\rm e}_{2}$)  &   5001.48  &    5.4344  &  -13.9923  &   25.1163  &  -14.3883  &    3.9862  &   -0.4843  &   -1.1799  &  0.120\\
2p3d($^3$F$^{\rm o}_{4}$) -- 2p3p($^3$D$^{\rm e}_{3}$)  &   5005.15  &    0.8597  &    2.2875  &   -0.4769  &    0.2119  &   -0.0007  &   -0.5229  &   -0.3551  &  0.097\\
2p3d($^3$F$^{\rm o}_{2}$) -- 2p3p($^3$D$^{\rm e}_{2}$)  &   5016.39  &    9.0157  &  -26.8968  &   41.0479  &  -25.4710  &    6.4845  &    0.0136  &   -1.5685  &  0.077\\
2p3d($^3$F$^{\rm o}_{3}$) -- 2p3p($^3$D$^{\rm e}_{3}$)  &   5025.66  &    9.3255  &  -19.1596  &   25.8965  &  -14.8728  &    3.6687  &    0.7105  &   -1.6874  &  0.033\\
2p3d($^3$F$^{\rm o}_{2}$) -- 2p3p($^3$D$^{\rm e}_{3}$)  &   5040.72  &  -10.6413  &    8.9049  &   -6.1303  &   -4.0371  &   -0.1665  &    1.9121  &   -2.7627  &  0.396\\
2p3d($^3$D$^{\rm o}_{2}$) -- 2p3p($^3$D$^{\rm e}_{1}$)  &   4774.24  &   -9.1418  &   11.6148  &   -9.7345  &    3.3344  &   -0.4313  &    1.4271  &   -1.0692  &  0.126\\
2p3d($^3$D$^{\rm o}_{1}$) -- 2p3p($^3$D$^{\rm e}_{1}$)  &   4779.72  &   10.2843  &  -18.3764  &   22.3938  &  -11.9676  &    2.8317  &    0.9316  &   -1.7007  &  0.029\\
\hline
\end{longtable}}

\longtab{13}{
\begin{longtable}{llrrrrrrr}
\caption[]{\label{fit4_lowTe} The same as Table~\ref{fit1_lowTe} but for $T_\mathrm{e} < 10,000$~K and $N_\mathrm{e} = 10^5$~cm$^{-3}$.}\\
\cline{1-9}\\
Transition & $\lambda$[\AA] & $a_0$ & $a_1$ & $a_2$ & $a_3$ & $a_4$ & $a_5$ & $\delta[\%]$ \\
\cline{1-9}\\
\endfirsthead

\caption{Continued.}\\
\cline{1-9}\\
Transition & $\lambda$[\AA] & $b_0$ & $a_1$ & $a_2$ & $a_3$ & $a_4$ & $a_5$ & $\delta[\%]$ \\
\cline{1-9}\\
\endhead
\hline
\endfoot

2p3d($^3$F$^{\rm o}_{2}$) -- 2p3p($^3$D$^{\rm e}_{1}$)  &   5001.14  &   13.4797  &   -7.9074  &    1.4674  &    0.2400  &   -0.1185  &    0.0110  &  0.204\\
2p3d($^3$F$^{\rm o}_{3}$) -- 2p3p($^3$D$^{\rm e}_{2}$)  &   5001.48  &   17.7235  &  -14.9425  &    6.1881  &   -1.2988  &    0.1276  &   -0.0045  &  0.127\\
2p3d($^3$F$^{\rm o}_{4}$) -- 2p3p($^3$D$^{\rm e}_{3}$)  &   5005.15  &   -1.2131  &   12.7460  &  -10.0136  &    3.4661  &   -0.5750  &    0.0370  &  0.460\\
2p3d($^3$F$^{\rm o}_{2}$) -- 2p3p($^3$D$^{\rm e}_{2}$)  &   5016.39  &   12.9566  &   -8.3346  &    1.7444  &    0.1515  &   -0.1045  &    0.0101  &  0.166\\
2p3d($^3$F$^{\rm o}_{3}$) -- 2p3p($^3$D$^{\rm e}_{3}$)  &   5025.66  &   16.6493  &  -14.7854  &    6.0663  &   -1.2523  &    0.1189  &   -0.0039  &  0.175\\
2p3d($^3$F$^{\rm o}_{2}$) -- 2p3p($^3$D$^{\rm e}_{3}$)  &   5040.72  &   11.8569  &   -9.2387  &    2.3497  &   -0.0490  &   -0.0717  &    0.0080  &  0.136\\
2p3d($^3$D$^{\rm o}_{2}$) -- 2p3p($^3$D$^{\rm e}_{1}$)  &   4774.24  &   10.3115  &   -8.0825  &    1.7211  &    0.1190  &   -0.0937  &    0.0091  &  0.122\\
2p3d($^3$D$^{\rm o}_{1}$) -- 2p3p($^3$D$^{\rm e}_{1}$)  &   4779.72  &   15.2449  &  -10.9597  &    3.0127  &   -0.1610  &   -0.0650  &    0.0081  &  0.151\\
\hline                                                                                                                                    
\end{longtable}}

\longtab{14}{
\begin{longtable}{llrrrrrrrr}
\caption[]{\label{fit4_highTe} The same as Table~\ref{fit1_lowTe} but for $10,000 \leq T_\mathrm{e} \leq 20,000$~K and $N_\mathrm{e} = 10^5$~cm$^{-3}$.}\\
\cline{1-10}\\
Transition & $\lambda$[\AA] & $b_0$ & $b_1$ & $b_2$ & $b_3$ & $b_4$ & $b_5$ & $b_6$ & $\delta[\%]$ \\
\cline{1-10}\\
\endfirsthead

\caption{Continued.}\\
\cline{1-10}\\
Transition & $\lambda$[\AA] & $b_0$ & $b_1$ & $b_2$ & $b_3$ & $b_4$ & $b_5$ & $b_6$ & $\delta[\%]$ \\
\cline{1-10}\\
\endhead
\hline
\endfoot

2p3d($^3$F$^{\rm o}_{2}$) -- 2p3p($^3$D$^{\rm e}_{1}$)  &   5001.14  &    8.6371  &  -28.7645  &   46.4861  &  -28.1871  &    7.5968  &   -0.7070  &   -1.2628  &  0.199\\
2p3d($^3$F$^{\rm o}_{3}$) -- 2p3p($^3$D$^{\rm e}_{2}$)  &   5001.48  &    9.7799  &  -15.5785  &   23.9781  &  -13.3185  &    3.6162  &    0.4909  &   -1.5046  &  0.149\\
2p3d($^3$F$^{\rm o}_{4}$) -- 2p3p($^3$D$^{\rm e}_{3}$)  &   5005.15  &    7.3195  &   -5.5325  &    6.9576  &   -2.5824  &    0.6538  &    0.5439  &   -1.2082  &  0.062\\
2p3d($^3$F$^{\rm o}_{2}$) -- 2p3p($^3$D$^{\rm e}_{2}$)  &   5016.39  &    9.6545  &  -29.9187  &   45.6077  &  -28.4729  &    7.2324  &   -0.0415  &   -1.5753  &  0.117\\
2p3d($^3$F$^{\rm o}_{3}$) -- 2p3p($^3$D$^{\rm e}_{3}$)  &   5025.66  &    9.3196  &  -19.8887  &   27.3066  &  -15.8535  &    3.9221  &    0.6653  &   -1.6868  &  0.048\\
2p3d($^3$F$^{\rm o}_{2}$) -- 2p3p($^3$D$^{\rm e}_{3}$)  &   5040.72  &   19.4889  &  -65.4463  &   75.0950  &  -35.3760  &   -0.1427  &   -0.8782  &   -2.0947  &  0.274\\
2p3d($^3$D$^{\rm o}_{2}$) -- 2p3p($^3$D$^{\rm e}_{1}$)  &   4774.24  &  -12.8694  &   17.4874  &  -14.5173  &    4.9124  &   -0.6436  &    1.6827  &   -1.3141  &  0.122\\
2p3d($^3$D$^{\rm o}_{1}$) -- 2p3p($^3$D$^{\rm e}_{1}$)  &   4779.72  &    9.8316  &  -18.5402  &   23.6227  &  -13.0596  &    3.1435  &    0.8475  &   -1.6969  &  0.040\\
\hline
\end{longtable}}

\begin{table*}
\caption{
Comparison of our direct recombination coefficients [cm$^{3}$ s$^{-1}$] to 
states of \ion{N}{$^+$} with those of Kisielius \& Storey \cite{KS02} 
and of Nahar \cite{Nahar95}. The comparison is for electron temperature 
$T_\mathrm{e} = 1,000$ and $10,000$~K, and electron density $N_\mathrm{e} 
= 10^4$~cm$^{-3}$.
}
\label{comparedir}
\centering
\begin{tabular}{lrrrrrr}
\cline{1-7} \\
& \multicolumn{3}{c}{1,000~K} & \multicolumn{3}{c}{10,000~K}\\
\cline{2-7} \\
State & Nahar & KS02 & Present &  Nahar & KS02 & Present \\
\cline{1-7} \\
2s$^2$2p$^2$~($^3$P)            & 2.44e$-$12 &  2.26e$-$12 &  2.28e$-$12 &  8.31e$-$13 &  7.69e$-$13 &  7.76e$-$13\\
2s$^2$2p$^2$~($^1$D)            & 1.28e$-$12 &  1.20e$-$12 &  1.19e$-$12 &  4.21e$-$13 &  3.92e$-$13 &  3.91e$-$13\\
2s$^2$2p$^2$~($^1$S)            & 2.55e$-$13 &  2.41e$-$13 &  2.39e$-$13 &  8.27e$-$14 &  7.75e$-$14 &  7.66e$-$14\\
\hline
\end{tabular}
\end{table*}

\end{document}